\documentclass[12pt]{article}
\usepackage{a4wide,graphicx,amsmath,amssymb,cite,longtable,multicol,array}

\newcommand{\lsim}{\raisebox{-0.13cm}{~\shortstack{$<$ \\[-0.07cm] $\sim$}}~}
\newcommand{\gsim}{\raisebox{-0.13cm}{~\shortstack{$>$ \\[-0.07cm] $\sim$}}~}
\newcommand{\Raw}{\Rightarrow}
\newcommand{\nn}{\nonumber}
\newcommand{\bea}{\begin{eqnarray}}
\newcommand{\eea}{\end{eqnarray}}
\newcommand{\ba}{\begin{array}}
\newcommand{\ea}{\end{array}}
\newcommand{\dis}{\displaystyle}
\newcommand{\ii}{{\rm i}}
\newcommand{\lag}{{\cal L}}

\renewcommand{\d}{{\rm d}}
\newcommand{\e}{{\rm e}}
\newcommand{\muegamma}{\mu\to\e\gamma}
\newcommand{\mueee}{\mu\to\e\e\bar{\e}}

\def\ket#1{\left| #1\right\rangle }
\def\bra#1{\left\langle #1\right| }
\def\slash#1{\setbox0=\hbox{$#1$}%
  \rlap{\ifdim\wd0>.7em\kern.22\wd0\else\kern.1\wd0\fi /}#1}
\def\bU{V_{\rm PMNS}}

\renewcommand{\arraystretch}{1.3}

\makeatletter
\@addtoreset{equation}{section}

\makeatother

\begin{document}

\hfill\begin{tabular}{r}
CAFPE-109/08 \\
UG-FT-239/08\\
\end{tabular}

\bigskip
\bigskip
\bigskip
\bigskip

\begin{center}

\begin{Large}
\textbf{Precise limits from lepton flavour violating processes \\[1ex]
        on the Littlest Higgs model with T-parity}
\end{Large}

\bigskip
\bigskip

\begin{large}
F.~del~\'Aguila, 
J.~I.~Illana
and
M.~D.~Jenkins
\end{large}

\bigskip

\begin{it}
CAFPE and Departamento de F{\'\i}sica Te\'orica y del Cosmos, \\[1ex]
    Universidad de Granada, E--18071 Granada, Spain
\end{it}

\bigskip

{\tt faguila@ugr.es}, {\tt jillana@ugr.es}, {\tt mjenk@ugr.es}
\end{center}

\bigskip
\bigskip

\begin{abstract}
We recalculate the leading one-loop contributions to 
$\muegamma$ and $\mueee$ 
in the Littlest Higgs model with T-parity, 
recovering previous results for the former. 
When all the Goldstone interactions are taken into account, 
the latter is also ultraviolet finite. 
The present experimental limits on these processes require 
a somewhat heavy effective scale $\sim 2.5$ TeV, or the flavour 
alignment of the Yukawa couplings of light and heavy leptons 
at the $\sim$ 10\% level, or the splitting of heavy lepton 
masses to a similar precision. Present limits on $\tau$ 
decays set no bounds on the corresponding parameters 
involving the $\tau$ lepton.
\end{abstract}

\thispagestyle{empty}

\newpage
\tableofcontents

\newpage

%%%%%%%%%%%%%%%%%%%%%%%%%%%%%%%%%%%%%%%%%%%%%%%%%%%%%%%%%%%%%%%%%%%%%%%%%%%%
\section{Introduction}

Little Higgs models \cite{ArkaniHamed:2001ca} offer an explanation to the little 
hierarchy between the Higgs mass $M_h$ assumed to be near 
the electroweak scale $v=246$~GeV and the new physics 
(NP) scale $f$, whose natural value is expected 
to be $\sim$ 1 TeV \cite{LH-reviews}. 
In contrast with supersymmetry, where the large one-loop 
Standard Model (SM) contributions to the Higgs mass are 
cancelled by the contributions from the corresponding 
supersymmetric partners with masses $\sim$ 1 TeV and 
spins differing by $\pm 1/2$ 
(see \cite{Wess:1992cp} and references therein), 
Little Higgs (LH) models stabilize $M_h$ by making the Higgs 
a pseudo-Goldstone boson of a broken global symmetry. 
The cancellation is in this case between particles 
with the same spin belonging to the same multiplets
of this approximate symmetry. 
Which of these SM extensions, if any, is at work will be 
hopefully established at the LHC \cite{LHC,Raidal:2008jk}. 

Supersymmetry is linearly realized in the minimal 
supersymmetric SM (MSSM). This and other simple supersymmetric 
extensions of the SM have other interesting phenomenological 
properties, like, for instance, the unification of gauge couplings 
at very high scales \cite{Dimopoulos:1981yj}, which is not the case 
for LH models. 
However, they have no built-in low energy mechanism 
to explain the observed fermion mass hierarchy or 
flavour conservation. 
As a matter of fact, it can be argued that supergravity 
is phenomenologically relevant 
\cite{Wess:1992cp,AlvarezGaume:1983gj}
because it can provide the necessary initial conditions to explain 
the precise fine-tuning required among the many new parameters 
of the MSSM, which otherwise would result in too large flavour
changing processes \cite{Dimopoulos:1981zb}.
This has been historically the problem of many SM extensions 
\cite{Dimopoulos:1980fj}. If the NP is near the TeV scale, it faces
in general the problem of naturally explaining why it is {\it aligned} 
with the SM Yukawa interactions, as experimentally required. 
Although the SM can not explain the large hierarchy 
between fermion masses, which by the way is several 
orders of magnitude more demanding than the little hierarchy, 
it naturally accommodates the absence of flavour 
changing neutral currents (FCNC) \cite{Glashow:1970gm}.
LH models are not designed to solve the flavour puzzle either, 
and one must expect stringent constraints on the new parameters 
involving the heavy sector. The study of 
FCNC processes in Littlest Higgs models 
has been addressed in the literature \cite{LH-FCNC}. 
In this paper we revise the calculation of the decay rates of the 
lepton flavour violating (LFV) processes 
$\muegamma$ \cite{Choudhury:2006sq,Blanke:2007db} 
and $\mueee$ \cite{Blanke:2007db}
in the Littlest Higgs model with T-parity (LHT) \cite{Cheng:2003ju}, 
obtaining an ultraviolet finite result also for the latter.\footnote{See also 
\cite{Goto:2008fj} which appeared when we were preparing this manuscript. 
There the cancellation of ultraviolet divergences in the LHT model is 
also shown, flavour violation in the quark sector is explored and the 
phenomenology of $K\to\pi\nu\bar\nu$ is analyzed.}
Indeed, when all Goldstone boson interactions of the 
new leptons are taken into account, the one-loop 
contributions to the amplitudes are well-defined \cite{Callan:1969sn,Georgi:1985kw},
scaling approximately in the two family case like 
$(v^2/f^2) \sin2\theta\ \delta$, 
where $\theta$ is a measure of the misalignment between the heavy 
and SM lepton Yukawa couplings and $\delta$ is the corresponding 
heavy lepton mass splitting. As a consequence, the present 
experimental limits require fine tuning the Yukawa couplings of 
the new heavy leptons up to 10\%, aligning them with their SM 
counterparts, or making the heavy masses quasi-degenerate. 
One might also rise the NP scale degrading the motivation of the 
LH scenario itself. 
In the general case with three families the new contributions must 
be tuned to a similar precision but the parameter dependence is 
more involved. 
The calculation also applies to $\tau$ decays, but the corresponding 
limits are not restrictive at present.
Moreover, it can be easily extended to $\mu-\e$ conversion in nuclei 
\cite{Hebert:2003kf}.
A complete phenomenological analysis comparing as well different LH 
models will be presented elsewhere.

In LH models the Higgs is a pseudo-Goldstone boson. 
Thus, $M_h$ is naturally small as long as the new scale $f$ 
is relatively low, because one expects that cancellations 
are only protected to one loop and for the dominant 
contributions. Hence, $4 \pi f$ can not be much larger 
than 10 TeV if we do not wish to invoke some fine tuning 
again. However, as the model introduces heavy particles 
the new one-loop contributions to electroweak observables 
may require rising $f$ significantly above 1 TeV 
in the absence of model dependent cancellations, 
in order to be consistent with present electroweak 
precision data (EWPD) \cite{EWPD}. 
The LHT is an economical realization of the LH scenario 
with the further virtue of keeping the new contributions 
to EWPD small. 
It incorporates a discrete symmetry under which the new particles are 
odd and the SM ones even. Then all vertices must have an even 
number of new particles, if any. Similarly to the R symmetry 
in supersymmetric models, the T symmetry allows us to weaken 
the experimental limit on the LH effective scale below 
the TeV \cite{Hubisz:2005tx}. 
This symmetry also makes stable the lightest T-odd particle, 
offering, like R-parity does in the supersymmetric case, an 
alternative candidate for cold dark matter \cite{Birkedal:2006fz}. 

Nevertheless, as already emphasized these models are not a priori 
designed to deal with the flavour problem. 
Therefore, it is important to investigate the constraints on 
the model parameters implied by the stringent experimental 
limits on FCNC. 
We follow an operational approach and calculate the leading 
contributions to 
$\muegamma$ and $\mueee$ in the LHT, 
recovering previous results for the former \cite{Choudhury:2006sq,Blanke:2007db} 
but an ultraviolet finite result for the latter. 
We focus on these processes because the lepton sector is free 
from large strong corrections, and the experimental limits are 
quite demanding. 
%We do not try to complete (improve) the model, in the sense 
%for instance of making sure (writing) that all new particles 
%including those not entering in our calculation are T odd \cite{???}. 
%Then, the Lagrangian will exhibit the continuous symmetries 
%explicitly.   
In Section \ref{LHT} we review the LHT model to introduce the 
notation and the Feynman rules needed. 
The one-loop amplitudes of the LFV 
processes  
$\muegamma$ and $\mueee$ 
are discussed in Section \ref{processes}. The 
calculation is straightforward but cumbersome, 
requiring a careful bookkeeping of the different terms. 
In Section \ref{limits} we present the numerical results 
discussing the dependence on the different parameters of the model.  
Finally, Section \ref{conclusions} is devoted to conclusions, 
where we also briefly comment on $\tau$ decays.

%%%%%%%%%%%%%%%%%%%%%%%%%%%%%%%%%%%%%%%%%%%%%%%%%%%%%%%%%%%%%%%%%%%%%%%%%%%%
\section{The Littlest Higgs model with T-parity}
\label{LHT}

\subsection{The Lagrangian}

The LHT 
is a non-linear $\sigma$ model based on the coset space 
SU(5)/SO(5), with the SU(5) global symmetry broken by the vacuum 
expectation value (VEV) of a $5 \times 5$ symmetric tensor,  
\bea
\Sigma_0=\left(\ba{ccc} {\bf 0}_{2\times2} & 0 & {\bf 1}_{2\times2} \\
                         0 & 1 &0 \\ 
                         {\bf 1}_{2\times2} & 0 & {\bf 0}_{2\times 2}\ea\right).
\label{Sigma}
\eea
The 10 unbroken generators $T^a$, 
which leave invariant $\Sigma_0$ 
and then satisfy $T^a \Sigma_0 + \Sigma_0 (T^a)^T = 0$,  
expand the SO(5) algebra; 
whereas the 14 broken generators $X^a$, 
which fulfill $X^a \Sigma_0 - \Sigma_0 (X^a)^T = 0$,  
expand the Goldstone fields $\Pi=\pi^a X^a$ parameterized as 
\bea
\Sigma(x)={\rm e}^{\ii\Pi/f}\Sigma_0{\rm e}^{\ii\Pi^T/f}={\rm e}^{2\ii\Pi/f}\Sigma_0, 
\label{Sigmaparameterization}
\eea
where $f$ is the effective NP scale.
Only the [SU(2)$\times$U(1)]$_1 \times$[SU(2)$\times$U(1)]$_2$ 
subgroup of the SU(5) global symmetry is gauged. It is generated by
\bea
Q_1^a=\frac{1}{2}\left(\ba{ccc}\sigma^a & 0 & 0 \\ 0 & 0 & 0 \\ 0 & 0 &  {\bf 0}_{2\times2} \ea\right),\quad
Y_1=\frac{1}{10}{\rm diag}(3,3,-2,-2,-2),
\\
Q_2^a=\frac{1}{2}\left(\ba{ccc}{\bf 0}_{2\times2} & 0 & 0 \\ 0 & 0 & 0 \\ 0 & 0 & -\sigma^{a*} \ea\right),\quad
Y_2=\frac{1}{10}{\rm diag}(2,2,2,-3,-3),
\eea
with $\sigma^a$ the three Pauli matrices. The VEV in Eq. (\ref{Sigma}) 
breaks this gauge group down to the SM gauge group SU(2)$_L\times$U(1)$_Y$, generated 
by the combinations $\{Q_1^a+Q_2^a,\ Y_1+Y_2\}\subset\{T^a\}$. The orthogonal 
combinations are a subset of the broken generators, 
$\{Q_1^a-Q_2^a,\ Y_1-Y_2\}\subset\{X^a\}$.
Thus, the Goldstone fields 
\bea
\Pi=\left(\ba{ccccc}
-\dis\frac{\omega^0}{2}-\frac{\eta}{\sqrt{20}} & -\dis\frac{\omega^+}{\sqrt{2}} & -\ii\dis\frac{\pi^+}{\sqrt{2}} & -\ii\Phi^{++} & -\ii\dis\frac{\Phi^+}{\sqrt{2}} \\
-\dis\frac{\omega^-}{\sqrt{2}} & \dis\frac{\omega^0}{2}-\frac{\eta}{\sqrt{20}} & \dis\frac{v+h+\ii\pi^0}{2} & -\ii\dis\frac{\Phi^+}{\sqrt{2}} & \dis\frac{-\ii\Phi^0+\Phi^P}{\sqrt{2}} \\
\ii\dis\frac{\pi^-}{\sqrt{2}} & \dis\frac{v+h-\ii\pi^0}{2} & \sqrt{\dis\frac{4}{5}}\eta & -\ii\dis\frac{\pi^+}{\sqrt{2}} &  \dis\frac{v+h+\ii\pi^0}{2} \\
\ii\Phi^{--} & \ii\dis\frac{\Phi^-}{\sqrt{2}} & \ii\dis\frac{\pi^-}{\sqrt{2}} & -\dis\frac{\omega^0}{2}-\frac{\eta}{\sqrt{20}} & -\dis\frac{\omega^-}{\sqrt{2}} \\
\ii\dis\frac{\Phi^-}{\sqrt{2}} & \dis\frac{\ii\Phi^0+\Phi^P}{\sqrt{2}} &  \dis\frac{v+h-\ii\pi^0}{2} & -\dis\frac{\omega^+}{\sqrt{2}} & \dis\frac{\omega^0}{2}-\frac{\eta}{\sqrt{20}}
\ea\right)
\label{goldstones}
\eea
decompose into the SM Higgs doublet $(-\ii\pi^+/\sqrt{2}, (v+h+\ii\pi^0)/2)^T$,
a complex SU(2)$_L$ triplet $\Phi$, and the 
longitudinal modes of the heavy gauge fields $\omega^\pm, \omega^0$ and $\eta$.\footnote{
In the following we use for the SM fields and couplings 
the conventions in Ref.~\cite{Denner:1991kt}. In particular,
$\phi^+=-\ii\pi^+$, $\phi^0=\pi^0$.}

As emphasized in the previous section, we can make the new 
contributions to electroweak precision observables small enough  
introducing a T-parity under which the SM particles are even 
and the new particles are odd. 
An obvious choice for the action of such T-parity on the gauge 
fields $G_i$ is the exchange of the gauge subgroups 
[SU(2)$\times$U(1)]$_1$ and [SU(2)$\times$U(1)]$_2$, 
\bea
G_1\stackrel{{\rm T}}{\longleftrightarrow} G_2.
\eea
Then, T invariance requires that the gauge couplings associated to 
both factors are equal. 
This leaves the following gauge Lagrangian unchanged,
\bea
\lag_G &=& \sum_{j=1}^2\left[-\frac{1}{2}{\rm Tr}
\left(\widetilde W_{j\mu\nu}\widetilde W_j^{\mu\nu}\right)
               -\frac{1}{4}B_{j\mu\nu}B_j^{\mu\nu}\right], 
\eea
where
\bea
\widetilde W_{j\mu} = W_{j\mu}^a Q_j^a, \quad
\widetilde W_{j\mu\nu}=
\partial_\mu\widetilde W_{j\nu}-\partial_\nu\widetilde W_{j\mu}
-\ii g\left[\widetilde W_{j\mu},\widetilde W_{j\nu}\right],\quad
B_{j\mu\nu}=\partial_\mu B_{j\nu}-\partial_\nu B_{j\mu}.
\eea
(Summation over index $a$, which runs on the corresponding 
SU(2) generators, is always assumed when repeated.) 
The T-even combinations multiplying the unbroken gauge generators 
correspond to the SM gauge bosons,  
\bea
W^\pm=\frac{1}{2}[(W_1^1+W_2^1)\mp\ii (W_1^2+W_2^2)],\quad
W^3=\frac{W_1^3+W_2^3}{\sqrt{2}},\quad
B=\frac{B_1+B_2}{\sqrt{2}}, 
\eea
whereas the T-odd combinations 
\bea
W_H^\pm=\frac{1}{2}[(W_1^1-W_2^1)\mp\ii (W_1^2-W_2^2)],\quad
W_H^3=\frac{W_1^3-W_2^3}{\sqrt{2}},\quad
B_H=\frac{B_1-B_2}{\sqrt{2}}, 
\eea
expand the heavy gauge sector.

In order to ensure that the SM Higgs doublet is T-even and 
the remaining Goldstone fields are T-odd, the 
T action on the scalar fields is defined as follows, 
\bea
\Pi\stackrel{{\rm T}}{\longrightarrow}-\Omega\Pi\Omega,\quad
\Omega={\rm diag}(-1,-1,1,-1,-1),  
\label{Pi}
\eea
where $\Omega$ is an element of the center of the gauge group,\footnote{
Note that we have reversed the sign of $\Omega$ as compared to the literature, 
to make it a group element.}
which commutes with $\Sigma_0$ but not with the full global symmetry. 
Then, 
\bea
\Sigma\stackrel{{\rm T}}{\longrightarrow}
\widetilde\Sigma=\Omega\Sigma_0\Sigma^\dagger\Sigma_0\Omega,
\label{SigmaT}
\eea
and the scalar Lagrangian
\bea
\lag_{S} &=& \frac{f^2}{8}{\rm Tr}\left[(D_\mu\Sigma)^\dagger(D^\mu\Sigma)\right],
\eea
with 
\bea
\label{derivative}
D_\mu\Sigma=\partial_\mu\Sigma-\sqrt{2}\ii\sum_{j=1}^2
\left[gW_{j\mu}^a(Q_j^a\Sigma+\Sigma Q_j^{aT})-g'B_{j\mu}(Y_j\Sigma+\Sigma Y_j^T)\right],
\eea
is also gauge and T-invariant. 

This discrete symmetry must be implemented in the fermion sector too. This
is less straightforward. 
In fact, there is no proposed model fulfilling the three desired 
conditions: to give masses to all (SM) fermions with Yukawa couplings, 
preserving a discrete symmetry under which all new particles are odd 
and the SM ones even, and keeping the full global symmetry before introducing 
the symmetry breaking. 
Although terms explicitly breaking the global symmetries at the 
Lagrangian level must manifest as {\it badly} behaved contributions 
to physical processes \cite{Georgi:1985kw}, 
this will not be our case since all the explicit couplings entering in the 
calculation we are interested in can be derived from Lagrangian terms 
which are symmetric. 
Following Refs.~\cite{Cheng:2004yc,Low:2004xc} 
we introduce two left-handed fermion doublets in incomplete SU(5) multiplets, 
one transforming just under SU(2)$_1$ and the other under SU(2)$_2$,
for each SM left-handed lepton doublet:
\bea
\Psi_1=\left(\ba{c} - \ii \sigma^2 l_{1L} \\ 0 \\ 0 \ea\right),\quad
\Psi_2=\left(\ba{c} 0 \\ 0 \\ - \ii \sigma^2 l_{2L} \ea\right), 
\label{Psimultiplets}
\eea
where $l_{iL}=\left(\ba{c} \nu_{iL} \\ \ell_{iL}\ea\right),\ i=1,2,$ and 
\bea
\Psi_1\longrightarrow V^*\Psi_1,\quad
\Psi_2\longrightarrow V\Psi_2,
\eea
under an SU(5) transformation $V$. 
We define the T-parity action on these fermions  
\bea
\Psi_1\stackrel{{\rm T}}{\longleftrightarrow}\Omega\Sigma_0\Psi_2.
\label{PsiT}
\eea
Then the usual T-even combination $\Psi_1+\Omega\Sigma_0\Psi_2$ remains light 
and is identified, up to the proper normalization, with the SM fermion doublet.
The T-odd combination $\Psi_1-\Omega\Sigma_0\Psi_2$ 
pairs with a right-handed doublet (eigenvector of T), in a 
complete SO(5) multiplet,
\bea
\Psi_R=\left(\ba{c} \cdot \\ \cdot \\ - \ii \sigma^2 l_{HR} \ea\right), \quad 
\Psi_R\stackrel{{\rm T}}{\longrightarrow}\Omega\Psi_R, \quad 
\Psi_R&\longrightarrow U\Psi_R, 
\eea
where $U$ is an SO(5) transformation defined below, to form a heavy Dirac doublet. 
With this aim in mind, a non-linear Yukawa Lagrangian is introduced,
\bea
\lag_{Y_H} = -\kappa f \left(\overline\Psi_2\xi+ 
\overline\Psi_1\Sigma_0\xi^\dagger\right)\Psi_R
+{\rm h.c.}\ ,
\label{mirror}
\eea
where $\xi={\rm e}^{\ii\Pi/f}$. This is indeed T-invariant, since 
Eq.~(\ref{Pi}) implies 
\bea
\xi \stackrel{{\rm T}}{\longrightarrow} \Omega\xi^\dagger\Omega,    
\eea
and invariant under global transformations,
\bea
\Sigma=\xi^2\Sigma_0\longrightarrow V\Sigma V^T \quad 
\Raw \quad \xi\longrightarrow V\xi U^\dagger \equiv U\xi\Sigma_0 V^T\Sigma_0 ,
\eea
where $V$ is the global SU(5) tranformation and $U$ a function of $V$ and $\Pi$ 
taking values in the Lie algebra of the unbroken SO(5). 
It must be noted that the gauge singlet 
$\chi_R$, completing the SO(5) representation 
\bea 
\Psi_R=\left(\ba{c} \tilde\psi_R \\ \chi_R \\ - \ii \sigma^2 l_{HR} \ea\right)  
\label{complete}
\eea
and assumed to be heavy, is T-even.\footnote{
If we had defined the T action on the fermions 
$\Psi_1\stackrel{{\rm T}}{\longleftrightarrow}-\Sigma_0\Psi_2$, 
$\Psi_R\stackrel{{\rm T}}{\longrightarrow}-\Psi_R$
and the Yukawa Lagrangian with $\Omega$'s, 
$\lag_{Y_H} = -\kappa f 
\left(\overline\Psi_2\xi+ \overline\Psi_1\Sigma_0
\Omega\xi^\dagger\Omega\right)\Psi_R + {\rm h.c.}$, 
all new fermions would be T-odd and the new Lagrangian 
invariant under the new T-parity \cite{Low:2004xc}, 
but not under the full global symmetry because $\Omega$ does not 
commute with SU(5) neither with SO(5), although it does commute 
with the gauge group. 
We must insist that the explicit couplings entering in our 
calculation are the same in both cases.} 
On the other hand, the extra doublet $\tilde\psi_R$, which is also assumed to be 
heavy enough to agree with EWPD, is T-odd as desired. 

We have just introduced all heavy fields we need. 
However, one important comment is in order. 
The Yukawa-type Lagrangian $\lag_{Y_H}$ fixes the transformation properties 
of the heavy fermions and then their gauge couplings, in particular 
the non-linear couplings of the right-handed heavy fermions \cite{Hubisz:2004ft},
\bea
\lag_F&=&
\ii\overline\Psi_1\gamma^\mu D_\mu^*\Psi_1 \ + \ 
\ii\overline\Psi_2\gamma^\mu D_\mu\Psi_2 \nn
\\
&+&\ii\overline\Psi_R\gamma^\mu\left(\partial_\mu+\frac{1}{2}\xi^\dagger 
(D_\mu\xi)+\frac{1}{2}\xi (\Sigma_0 D_\mu^* \Sigma_0 \xi^\dagger)\right)\Psi_R
\label{RHkineticterm}
\eea
with
\bea
D_\mu&=&\partial_\mu-\sqrt{2}\ii g(W_{1\mu}^a Q_1^a+W_{2\mu}^a Q_2^a)
+\sqrt{2}\ii g'\left(Y_1 B_{1\mu}+Y_2 B_{2\mu}\right). 
\eea
The Lagrangian of Eq. (\ref{RHkineticterm}) includes the proper $O(v^2/f^2)$ 
couplings to Goldstone fields, 
absent in \cite{Blanke:2007db,Blanke:2006eb}, that render the one-loop amplitudes ultraviolet finite. 
Besides, in order to assign the proper SM hypercharge $y=-1$ to the charged 
right-handed leptons $\ell_R$, which are SU(5) singlets and T-even, one can 
enlarge SU(5) with two extra U(1) groups, since otherwise their hypercharge 
would be zero. Then, the corresponding gauge and T invariant Lagrangian reads
\bea
\lag'_F=\ii\overline \ell_R\gamma^\mu (\partial_\mu+\ii g' y B_\mu)\ell_R .
\eea

For the lepton sector and the calculation we are interested in 
these are all the necessary Lagrangian terms. However, in order to define what 
a muon or an electron is, we have to diagonalise the 
mass matrix $(M_\ell)_{ij} = (\lambda_\ell)_{ij} v$ 
in the corresponding Yukawa Lagrangian $\lag_Y$ which we assume to have 
all required properties \cite{Hubisz:2004ft,Chen:2006cs} \footnote{
Right-handed leptons, as the other right-handed SM fermions, 
are usually taken to be singlets under the 
non-abelian symmetries, transforming only 
under the gauge abelian subgroup. 
We must note that this may be a too strong assumption. 
If we want to couple them to their left-handed counterpart, one
may be inspired by the following observation. There is 
only one SU(5) singlet in the decomposition of the product of two $\Sigma$'s 
and one left-handed fermion multiplet,
$
\sum_{\alpha_i=1}^5 
\epsilon ^{\alpha_1\alpha_2\alpha_3\alpha_4\alpha_5}
[(\Sigma)_{\alpha_1 \alpha_2}(\Sigma)_{\alpha_3 \alpha_4} \Psi_{2 \alpha_5} 
+ (\Sigma^\dagger)_{\alpha_1 \alpha_2}(\Sigma^\dagger)_{\alpha_3 \alpha_4} 
\Psi_{1 \alpha_5}],   
%\label{LY5}
$ 
where $\epsilon ^{\alpha_1\alpha_2\alpha_3\alpha_4\alpha_5}$ 
is the totally antisymmetric tensor and the second term is the 
T transformed of the first one. Alternatively, one could multiply
three $\Sigma$'s and the other left-handed fermion multiplet, 
$
\sum_{\alpha_i=1}^5 
\epsilon ^{\alpha_1\alpha_2\alpha_3\alpha_4\alpha_5}\delta^{\alpha_6\alpha_7}
[(\Sigma)_{\alpha_1 \alpha_2}(\Sigma)_{\alpha_3 \alpha_4} (\Sigma)_{\alpha_5 \alpha_6} 
\Psi_{1 \alpha_7} 
+ (\Sigma^\dagger)_{\alpha_1 \alpha_2}(\Sigma^\dagger)_{\alpha_3 \alpha_4} 
(\Sigma^\dagger)_{\alpha_5 \alpha_6} \Psi_{2 \alpha_7}],   
%\label{LY7}
$
with $\delta^{\alpha_6\alpha_7}$ the Kronecker delta.
In both cases, we get the wrong Higgs coupling. 
This is so because this product is an SU(5) singlet and 
then the Higgs coupling reads 
$i\pi^+ l^- + (v + h + i\pi^0) \nu / \sqrt 2$. 
(In these expressions there are neither $\Omega$'s nor $\Sigma_0$'s  
because the determinant of $\Omega$ is 1 and 
$
\epsilon ^{\alpha_1\alpha_2\alpha_3\alpha_4\alpha_5}
(\Sigma_0)_{\alpha_1 \beta_1}(\Sigma_0)_{\alpha_2 \beta_2}
(\Sigma_0)_{\alpha_3 \beta_3}(\Sigma_0)_{\alpha_4 \beta_4}
(\Sigma_0)_{\alpha_5 \beta_5} = 
\epsilon ^{\beta_1\beta_2\beta_3\beta_4\beta_5}. 
$)
Then, getting the correct coupling 
$(v + h -i\pi^0) l^- / \sqrt 2 + i\pi^- \nu$ 
requires the explicit breaking of SU(5). 
If $\xi$ is introduced in the game, one eventually has to break SO(5) as well.} 
(and also to include light neutrino masses). 
This gives to leading order a mass term for the charged leptons 
$m_{\ell^i} \overline \ell^i_L \ell^i_R +\mbox{h.c.}$, with 
\bea
m_{\ell_{i'}} \delta_{i'j'} = 
(V_L^{\ell \dagger})_{i'i} (\lambda_\ell)_{ij} (V_R^\ell)_{jj'}\ v
\label{SMmixing}
\eea
and $V^\ell_{L,R}$ two unitary matrices.\footnote{
We denote the mass eigenstates with primes when necessary to 
distinguish them from the current eigenstates.}

Finally, in order to perform the calculation in the mass eigenstate basis we 
have to diagonalise the full Lagrangian 
\bea
\lag = 
\lag_{G} + \lag_{S} + \lag_{Y_H} + \lag_{F}  + \lag_F' + \lag_Y,
\eea 
and reexpress it in the mass eigenstate basis. The corresponding masses 
and eigenvectors up to order $v^2/f^2$ are given in Appendix~\ref{apppf}. 
The Feynman rules are collected in Appendix~\ref{appfr}. They are obtained 
expanding $\lag$ to the required order. 
The coupling overlooked in \cite{Blanke:2007db} is the $v^2/f^2$ correction 
to the right-handed coupling $g_R$ of the $Z\bar \nu_H^i\nu_H^j$ 
vertex, resulting from the expansion of the last two 
terms of $\lag_F$ in Eq.~(\ref{RHkineticterm}). 
 
%%%%%%%%%%%%%%%%%%%%%%%%%%%%%%%%%%%%%%%%%%%%%%%%%%%%%%%%%%%%%%%%%%%%%%%%%%%%
\subsection{Flavour mixing \label{flavour}}

The new contributions to charged LFV processes must be proportional to the ratio of the 
electroweak and the LHT breaking scales $v^2/f^2$ and to a combination 
of the matrix elements describing the misalignment of the heavy and 
charged lepton Yukawa couplings. 
Let us then set our conventions for the description of the heavy-light mixing
relevant to our analysis, and in particular to the Feynman rules discussed above and collected in Appendix~\ref{appfr}. The SM interaction and mass eigenstates are related by  
the unitary matrices in Eq.~(\ref{SMmixing}), 
\bea
\ell_L= V_L^\ell \ell'_L,\quad \ell_R= V_R^\ell \ell'_R.  
\eea 
Then the SM charged current Lagrangian reads 
\bea
\lag^{\rm SM}_{CC} = 
- \frac{g}{\sqrt{2}} \overline \nu_L \slash{W}^\dagger \ell_L + {\rm h.c.} = 
- \frac{g}{\sqrt{2}} \overline \nu'_L 
V_L^{\nu \dagger} V_L^\ell \slash{W}^\dagger \ell'_L + {\rm h.c.},
\eea
where we have also introduced the corresponding rotation for 
the neutrinos. 
Thus, only the combination 
$V_{\rm PMNS}^\dagger = V_L^{\nu \dagger} V_L^\ell$ is 
observable. 
It must be noted, however, that the neutrino contributions 
to LFV processes are negligible in the SM because so are 
their masses. Hence, $V_L^\nu$ can be assumed to be unity. 
Similarly we can also diagonalise the heavy Yukawa couplings 
in Eq. (\ref{mirror}),
\bea
m_{l_H^{i'}} \delta_{i'j'} = 
(V_L^{H \dagger})_{i'i} \kappa_{ij} (V_R^H)_{jj'}\ \sqrt 2 f, 
\label{Hmixing}
\eea
where $V_L^H$ acts on the left-handed fields and $V_R^H$ acts on the 
right-handed fields. 
Note that there is no distinction between up- and down-type leptons. 
The T-odd gauge boson interactions arising from the corresponding 
kinetic terms for left-handed leptons in  Eq.~(\ref{RHkineticterm})
are proportional to 
\bea
\overline l_{L -}\slash{G}_- l_{L +} + {\rm h.c.}
= \overline l_{H L} V^{H \dagger}_L \slash{G}_H \left(\ba{c} V^\nu_L \nu_L \\ 
V^\ell_L \ell_L \ea\right) + {\rm h.c.}
\eea
where $G_-$ and $l_{L -}$ are the heavy, T-odd gauge bosons and fermions 
and $l_{L +}$ are the SM, T-even fermions in the interaction basis, whereas 
$G_H = A_H,Z_H,W_H$; $l_H=(\nu_H, \ell_H)^T$; and 
$\nu_L$ and $\ell_L$ are the corresponding mass eigenstates.
Then, in analogy with the PMNS matrix, the observable rotations are now
\bea
V_{H\nu}\equiv V_L^{H \dagger} V_L^\nu,\quad
V_{H\ell}\equiv V_L^{H \dagger} V_L^\ell.
\eea
Note that both matrices are related, $V_{H \nu}^\dagger V_{H \ell}=V_{\rm PMNS}^\dagger$ 
\cite{Hubisz:2005bd}, but this relation can not be tested unless $V_{H \nu}$ can be 
measured. The new contributions to the LFV amplitudes describing a muon decay 
to an electron are then proportional to $V_{H\ell}^{ie*} V_{H\ell}^{i\mu}$, with 
$i$ counting the heavy lepton doublets.

%%%%%%%%%%%%%%%%%%%%%%%%%%%%%%%%%%%%%%%%%%%%%%%%%%%%%%%%%%%%%%%%%%%%%%%%%%%%
\section{New contributions to LFV processes 
\label{processes}}

As noted above, the SM contributions to the LFV processes 
$\muegamma$ and $\mueee$ are 
negligible for they are proportional to the observed neutrino 
masses. 
On the other hand the new LHT contributions can be a priori large. 
In particular, one expects that the dominant contributions come 
from the exchange of the new vector bosons and heavy fermions 
required to realise the discrete symmetry T.\footnote{
The addition of new vector-like leptons in general imply
large FCNC already at tree level \cite{Glashow:1970gm,delAguila:1982yu}, 
and stringent constraints from EWPD and LFV processes 
\cite{delAguila:2008pw}. In the LHT
they are absent because T-parity forbids the coupling of a
SM gauge boson to one light and one heavy fermion. Analogously,
the presence of heavy scalar triplets with hypercharge 1 in
general allows for their direct coupling to two (SM) lepton
doublets (for a review and further references see 
\cite{Raidal:2008jk,delAguila:2008cj}).
This is also absent in the LHT because the triplet $\Phi$ in
Eq. (\ref{goldstones}) is T-odd and the SM leptons are T-even.}
Here we study both processes in turn. 

%%%%%%%%%%%%%%%  FIGURE %%%%%%%%%%%%%%%%%%%%%%%%%
\begin{figure}[htb]
\centerline{\includegraphics[scale=0.8]{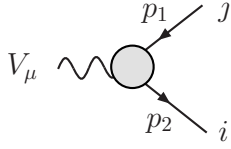}
%\quad\raisebox{11mm}{$\ii\Gamma^\mu(p_1,p_2)$}
}
\caption{Effective vector-fermion vertex.
\label{vff-0}}
\end{figure}
%%%%%%%%%%%%%%%%%%%%%%%%%%%%%%%%%%%%%%%%%%%%%%%%%

The amplitude $\muegamma$ is proportional to the vertex in 
Fig.~\ref{vff-0}, whose most general structure for on-shell fermions 
$f_{i,j}$ can be written in terms of six form factors:
\bea
\ii\Gamma^\mu(p_1,p_2) &=& \ii e\big[
	 \gamma^\mu(F_L^VP_L+F_R^VP_R)
	+(\ii F_{M}^V+F_{E}^V\gamma_5)\sigma^{\mu\nu}Q_\nu 
	+(\ii F_S^V+F_P^V\gamma_5)Q^\mu
      \big],\quad\quad
\eea
with $P_{R,L}=\frac{1}{2}(1\pm\gamma_5)$ and $Q=p_2-p_1$ 
the vector boson momentum entering into the vertex. 
If the vector boson $V$ is a photon, the U(1) gauge symmetry  
is unbroken and current conservation implies 
\bea
(m_i-m_j)(F_L^\gamma+F_R^\gamma) + 2\ii Q^2 F_S^\gamma &=&0,
\label{cc1}
\\
-(m_i+m_j)(F_L^\gamma-F_R^\gamma)+ 2Q^2 F_P^\gamma &=&0. 
\label{cc2}
\eea
Hence, the LFV process $f_j\to f_i\gamma$ 
with $i\ne j$ where the photon is on-shell ($Q^2=0$) is completely described 
by a dipole transition. Indeed, according to Eqs.~(\ref{cc1},\ref{cc2})  
$F_L^\gamma=F_R^\gamma=0$ for on-shell photons, 
while the form factors $F_{S,P}^\gamma$ do not contribute to the amplitude 
because real photons are transverse. 
Then, the total width for $\ell_j\to\ell_i\gamma$ is 
given by 
\cite{delAguila:1982yu,Hisano:1995cp,Illana:2000ic,Illana:2002tg,Arganda:2005ji}
\bea
\Gamma(\ell_j\to\ell_i\gamma)=\frac{\alpha}{2}m_{\ell_j}^3
\left(|F_M^\gamma|^2+|F_E^\gamma|^2\right). 
\eea

%%%%%%%%%%%%%%%  FIGURE %%%%%%%%%%%%%%%%%%%%%%%%%
\begin{figure}[htb]
\centerline{
\includegraphics[scale=0.8]{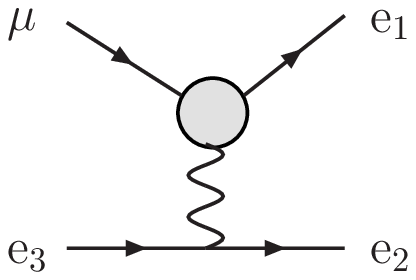}
\includegraphics[scale=0.8]{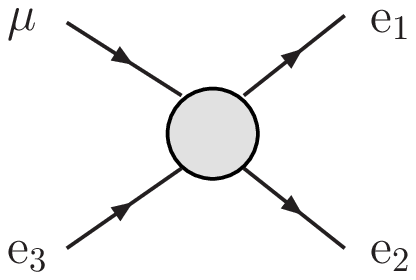}
}
\caption{Generic penguin and box diagrams for $\mu\to\e_1\e_2\bar\e_3$.
Crossed diagrams with $\e_1$ and $\e_2$ exchanged must be added.
\label{penguin-box}}
\end{figure}
%%%%%%%%%%%%%%%%%%%%%%%%%%%%%%%%%%%%%%%%%%%%%%%%%
On the other hand, two types of diagrams contribute to $\mueee$
(see Fig.~\ref{penguin-box}). Now, in the diagrams of the first type (penguins) 
the exchanged gauge boson $V$ can be a $\gamma$ or a Z but not a heavy vector 
boson for the coupling is forbidden by T-parity. (Higgs-penguins are neglected.) 
$F_L^V$ and $F_R^V$ do not vanish in penguins. 
In fact, for $\gamma$ these form factors are proportional to $Q^2$, 
as we have explicitly checked.
Besides, as the gauge boson couples to two on-shell electrons, 
the contributions from $F_{S,P}^V$ are irrelevant 
for they multiply the electron mass. 
The amplitude for this process also receives 
contributions from box diagrams.  
The total amplitude for $\mu(p)\to\e(p_1)\ \e(p_2)\ \bar\e(p_3)$ 
can be then written \cite{Hisano:1995cp}
\bea
{\cal M}&=&{\cal M}_{\gamma{\rm -penguin}}+{\cal M}_{Z{\rm -penguin}}
+{\cal M}_{\rm box},
\eea
with 
\bea
{\cal M}_{\gamma{\rm -penguin}}&=&
\frac{e^2}{Q^2}
\bar u(p_1)\left[Q^2\gamma^\mu(A_1^L P_L+A_1^R P_R)+m_\mu\ii \sigma^{\mu\nu}Q_\nu(A_2^LP_L+A_2^RP_R)\right]u(p)\
 \nn\\ && \times\bar u(p_2)\gamma_\mu v(p_3)- (p_1\leftrightarrow p_2),
\eea
\bea
{\cal M}_{Z{\rm -penguin}}&=&
\frac{e^2}{M_Z^2}
\bar u(p_1)\left[\gamma^\mu(F_L P_L+F_R P_R)\right]u(p)\
\bar u(p_2)\left[\gamma_\mu(Z_L^e P_L+Z_R^e P_R)\right] v(p_3) 
\nn\\ && - (p_1\leftrightarrow p_2),
\label{zpenguin}
\eea
\bea
{\cal M}_{\rm box}&=&
\quad e^2 B_1^L\left[\bar u(p_1)\gamma^\mu P_L u(p)\right]
            \left[\bar u(p_2)\gamma_\mu P_L v(p_3)\right] \nn\\
&&+e^2 B_1^R\left[\bar u(p_1)\gamma^\mu P_R u(p)\right]
            \left[\bar u(p_2)\gamma_\mu P_R v(p_3)\right] \nn\\
&&+e^2 B_2^L\left\{\left[\bar u(p_1)\gamma^\mu P_L u(p)\right]
            \left[\bar u(p_2)\gamma_\mu P_R v(p_3)\right]
             - (p_1\leftrightarrow p_2) \right\} \nn\\
&&+e^2 B_2^R\left\{\left[\bar u(p_1)\gamma^\mu P_R u(p)\right]
            \left[\bar u(p_2)\gamma_\mu P_L v(p_3)\right]
             - (p_1\leftrightarrow p_2) \right\} \nn\\
&&+e^2 B_3^L\left\{\left[\bar u(p_1) P_L u(p)\right]
            \left[\bar u(p_2) P_L v(p_3)\right]
             - (p_1\leftrightarrow p_2) \right\} \nn\\
&&+e^2 B_3^R\left\{\left[\bar u(p_1) P_R u(p)\right]
            \left[\bar u(p_2) P_R v(p_3)\right]
             - (p_1\leftrightarrow p_2) \right\} \nn\\
&&+e^2 B_4^L\left\{\left[\bar u(p_1) \sigma^{\mu\nu}P_L u(p)\right]
            \left[\bar u(p_2)\sigma_{\mu\nu} P_L v(p_3)\right]
             - (p_1\leftrightarrow p_2) \right\} \nn\\
&&+e^2 B_4^R\left\{\left[\bar u(p_1) \sigma^{\mu\nu}P_R u(p)\right]
            \left[\bar u(p_2)\sigma_{\mu\nu} P_R v(p_3)\right]
             - (p_1\leftrightarrow p_2) \right\} .
\label{MB}
\eea
We have defined new vertex form factors in the penguin amplitudes    
\bea
&&A_1^L=F_L^\gamma/Q^2, \
  A_1^R=F_R^\gamma/Q^2, \
  A_2^L=(F_M^\gamma+\ii F_E^\gamma)/m_\mu, \
  A_2^R=(F_M^\gamma-\ii F_E^\gamma)/m_\mu, \quad\quad\nn\\
&&F_L=-F_L^Z, \quad
  F_R=-F_R^Z ,
\eea
and used that $Q^2 \ll M_Z^2$ in Eq.~(\ref{zpenguin}). 
$Z_{L,R}^e$ are the corresponding Z couplings to the electron in the SM
(see Appendix~\ref{appfr}). 
The dipole form factors $F_{M,E}^Z$ are dropped from the amplitude 
because their contributions are effectively suppressed 
by a factor $m_\mu^2/M_{W_H}^2$. 
The total width can then be written as \cite{Hisano:1995cp,Arganda:2005ji}:
{\allowdisplaybreaks
\begin{align}
\Gamma(&\mueee)=\frac{\alpha^2m_\mu^5}{32\pi}\bigg[
|A_1^L|^2+|A_1^R|^2-2(A_1^LA_2^{R*}+A_2^LA_1^{R*}+{\rm h.c.}) 
\nn\\ & 
+(|A_2^L|^2+|A_2^R|^2)\left(\frac{16}{3}\ln\frac{m_\mu}{m_e}-\frac{22}{3}\right)
+\frac{1}{6}(|B_1^L|^2+|B_1^R|^2)
+\frac{1}{3}(|B_2^L|^2+|B_2^R|^2) \nn\\ & 
+\frac{1}{24}(|B_3^L|^2+|B_3^R|^2)
+6(|B_4^L|^2+|B_4^R|^2)
-\frac{1}{2}(B_3^LB_4^{L*}+B_3^RB_4^{R*}+{\rm h.c.}) \nn\\ & +\frac{1}{3}(A_1^LB_1^{L*}+A_1^RB_1^{R*}+A_1^LB_2^{L*}+A_1^RB_2^{R*}+{\rm h.c.}) \nn\\ & -\frac{2}{3}(A_2^RB_1^{L*}+A_2^LB_1^{R*}+A_2^LB_2^{R*}+A_2^RB_2^{L*}+{\rm h.c.}) \nn\\ & 
+\frac{1}{3}\big\{2(|F_{LL}|^2+|F_{RR}|^2)+|F_{LR}|^2+|F_{RL}|^2 \nn\\ &
+(B_1^LF_{LL}^*+B_1^RF_{RR}^*+B_2^LF_{LR}^*+B_2^RF_{RL}^*+{\rm h.c.})
+2(A_1^LF_{LL}^*+A_1^RF_{RR}^*+{\rm h.c.})  \nn\\ &
+(A_1^LF_{LR}^*+A_1^RF_{RL}^*+{\rm h.c.})
-4(A_2^RF_{LL}^*+A_2^LF_{RR}^*+{\rm h.c.}) \nn\\ &
-2(A_2^LF_{RL}^*+A_2^RF_{LR}^*+{\rm h.c.})\big\}\bigg],
\label{mueee}
\end{align}}
where
\bea
F_{LL}=\frac{F_LZ_L^e}{M_Z^2},\quad
F_{RR}=\frac{F_RZ_R^e}{M_Z^2},\quad
F_{LR}=\frac{F_LZ_R^e}{M_Z^2},\quad
F_{RL}=\frac{F_RZ_L^e}{M_Z^2}.
\eea
Note that the amplitude for the Z-penguin could have been cast into the box structure 
replacing 
\bea
B_1^L&\to& B_1^L + 2F_{LL} , \\
B_1^R&\to& B_1^R + 2F_{RR} ,\\
B_2^L&\to& B_2^L + F_{LR} ,\\
B_2^R&\to& B_2^R + F_{RL} .
\eea

The branching ratios for both types of processes are obtained dividing by 
the SM decay width
\bea
\Gamma(\ell_j\to\ell_i\nu_j\bar\nu_i)&=&\frac{G_F^2 m_{\ell_j}^5}{192\pi^3},\quad
G_F=\frac{\pi\alpha}{\sqrt{2}s_W^2M_W^2}. 
\eea
For $\tau$ decays the SM branching ratio must be corrected multiplying by 0.17 
to take into account other possible decay channels. 

%%%%%%%%%%%%%%%%%%%%%%%%%%%%%%%%%%%%%%%%%%%%%%%%%%%%%%%%%%%%%%%%%%%%%%%%%%%%
\subsection{$\muegamma$}

%%%%%%%%%%%%%%%  FIGURE %%%%%%%%%%%%%%%%%%%%%%%%%
\begin{figure}
\centerline{\small\begin{tabular}{ccc}
\includegraphics[scale=0.7]{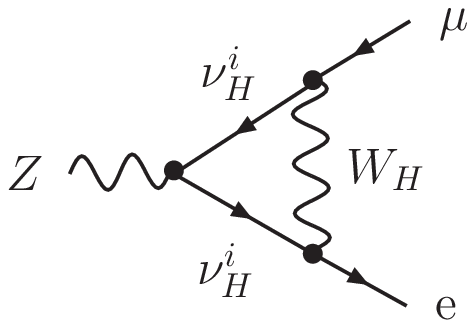} &
&
\includegraphics[scale=0.7]{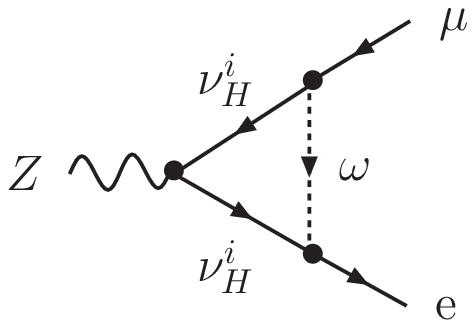} \\
\includegraphics[scale=0.7]{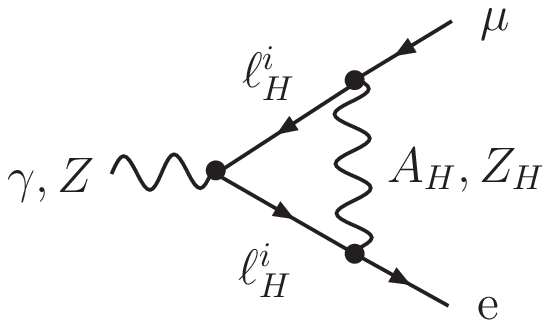} &
\includegraphics[scale=0.7]{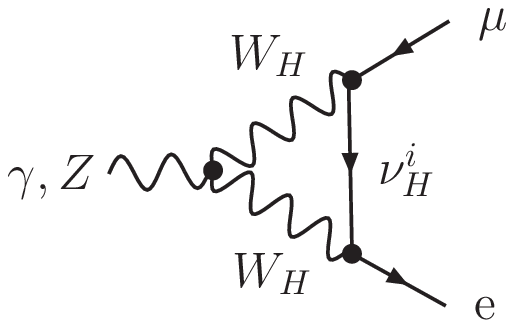} &
\includegraphics[scale=0.7]{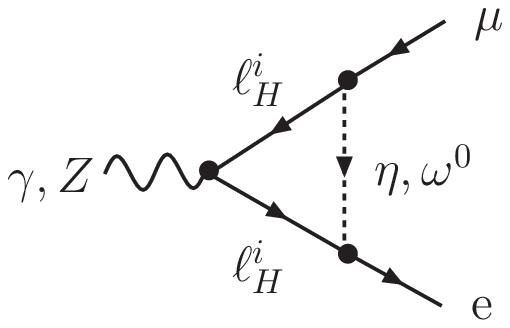} \\
I & II & III \\
\includegraphics[scale=0.7]{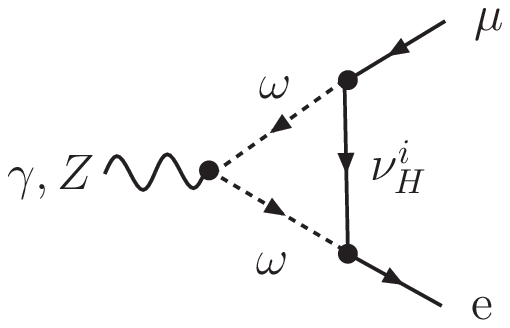} &
\includegraphics[scale=0.7]{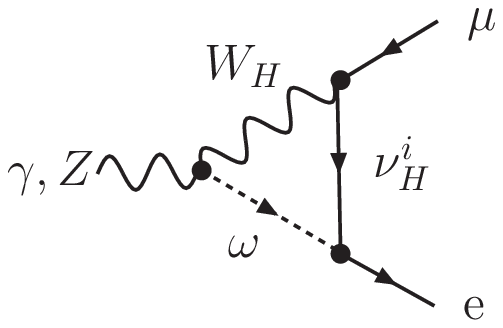} &
\includegraphics[scale=0.7]{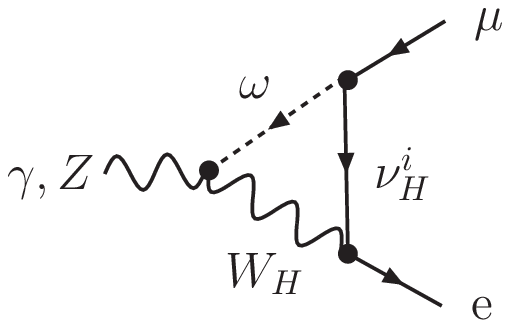} \\
IV & V & VI
\end{tabular}}

\centerline{
\includegraphics[scale=0.7]{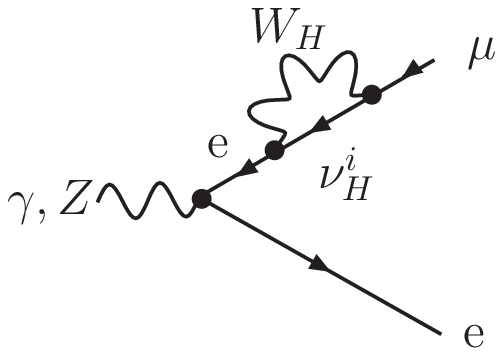}
\includegraphics[scale=0.7]{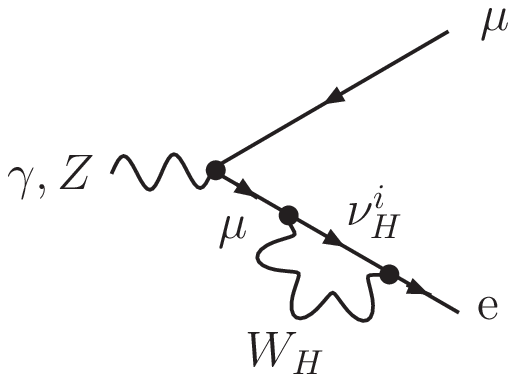}
\includegraphics[scale=0.7]{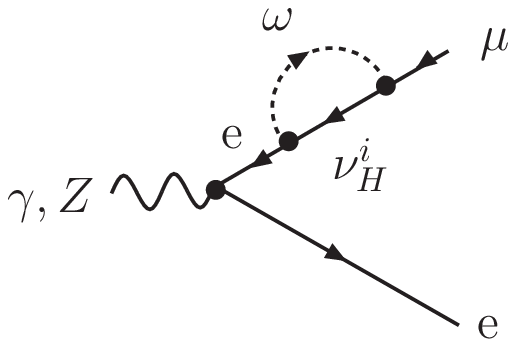}
\includegraphics[scale=0.7]{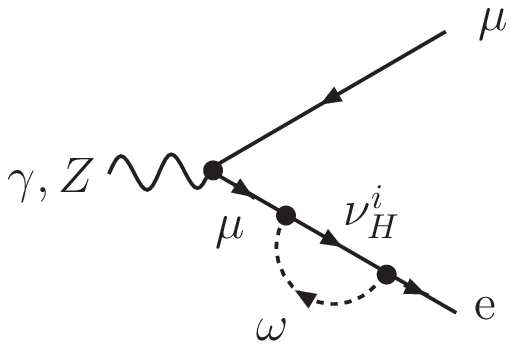}
}

\centerline{
\includegraphics[scale=0.7]{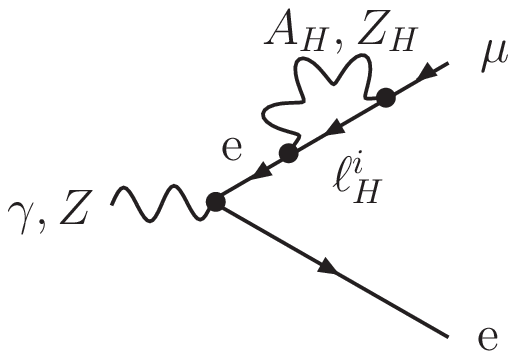}
\includegraphics[scale=0.7]{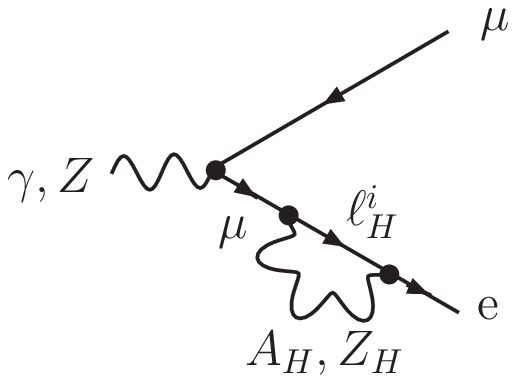}
\includegraphics[scale=0.7]{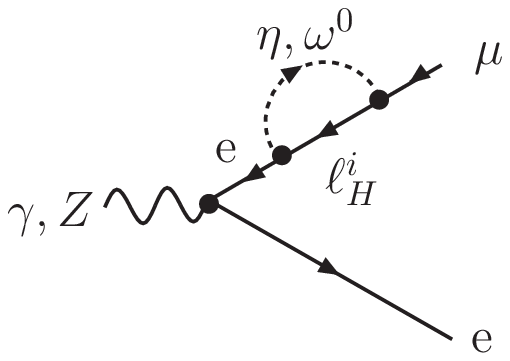}
\includegraphics[scale=0.7]{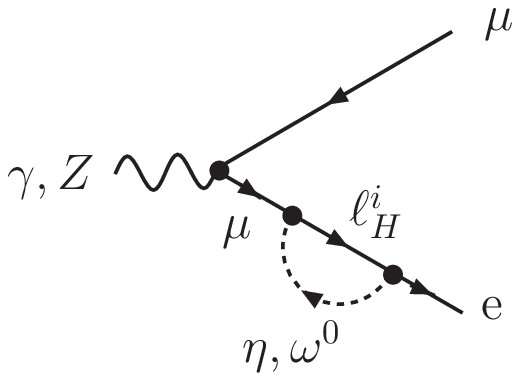}
}
\caption{New one-loop diagrams contributing to 
$V\mu\e$ in the LHT model.\label{Vff-LH}}
\end{figure}
%%%%%%%%%%%%%%%%%%%%%%%%%%%%%%%%%%%%%%%%%%%%%%%%%
Let us now summarize the calculation for $\muegamma$. 
The new one-loop Feynman diagrams contributing to the $V\mu\e$ vertex 
in the LHT model in the 't Hooft-Feynman gauge are listed in Fig.~\ref{Vff-LH}. 
They are classified in six topology classes. 
As explained above, in the decay $\muegamma$ the photon is on-shell 
and then only the dipole form factors $F_{M,E}^\gamma$ contribute. 
They are proportional to the muon mass,  
reflecting the chirality flip character of the dipole transition. 
(The electron mass is neglected.) 
We separate the contributions exchanging $W_H, Z_H$ and $A_H$, 
expressing the results in terms of standard loop integrals 
(Appendix~\ref{apploop}). 

\subsubsection*{Diagrams exchanging $W_H$}

Taking $M_1=M_{W_H}$ and $M_2=m_{\nu_H^i}$ and introducing the mass ratio
\bea
y_i=\frac{m_{H i}^2}{M_{W_H}^2},
\eea
with $m_{Hi}\equiv m_{\ell_H^i}\simeq m_{\nu_H^i}$, 
we find the following contributions from diagrams exchanging $W_H$ 
(see Fig.~\ref{Vff-LH}):
\bea
\mbox{II:}&&
F_M^\gamma|_{W_H}=-\ii F_E^\gamma|_{W_H} =
-\frac{\alpha_W}{16\pi}m_\mu\sum_iV_{H\ell}^{ie*}V_{H\ell}^{i\mu} \ \left[3\overline{C}_{11}-\overline{C}_1\right],
\\
\mbox{IV:}&&
F_M^\gamma|_{W_H}=-\ii F_E^\gamma|_{W_H} =
-\frac{\alpha_W}{16\pi}m_\mu\sum_iV_{H\ell}^{ie*}V_{H\ell}^{i\mu}\ y_i \ \left[\overline{C}_0+3\overline{C}_1+\frac{3}{2}\overline{C}_{11}\right],
\\
\mbox{V:}&&
F_M^\gamma|_{W_H}=-\ii F_E^\gamma|_{W_H} = 0 ,
\\
\mbox{VI:}&&
F_M^\gamma|_{W_H}=-\ii F_E^\gamma|_{W_H} =
\frac{\alpha_W}{16\pi}m_\mu\sum_iV_{H\ell}^{ie*}V_{H\ell}^{i\mu}\  \overline{C}_1 ,
\\
\mbox{Total:}&&
F_M^\gamma|_{W_H}=-\ii F_E^\gamma|_{W_H} =
\frac{\alpha_W}{16\pi}\frac{m_\mu}{M_{W_H}^2}\sum_iV_{H\ell}^{ie*}V_{H\ell}^{i\mu}\ 
F_W(y_i) ,
\label{fmW}
\eea
where $\alpha_W\equiv\alpha/s_W^2$ and
\bea
F_W(x) &=&M_1^2\left[
2\overline{C}_1-3\overline{C}_{11}-x\left(\overline{C}_0+3\overline{C}_1+\frac{3}{2}\overline{C}_{11}\right)\right]
\nonumber \\
&=& 
\frac{5}{6}-\frac{3x-15x^2-6x^3}{12(1-x)^3}+\frac{3x^3}{2(1-x)^4}\ln x .
\label{F_W}
\eea
The constant term drops from the amplitude due to the unitarity of the mixing matrix.
This result is in agreement with \cite{Cheng:1985bj,Lavoura:2003xp,Blanke:2007db}.

It may be worth to note that 
these contributions are completely analogous to those of the SM with massive neutrinos, replacing $W_H$ by $W$, $\nu_H^i$ by $\nu_i$ and $V_{H\ell}$ by $V_{\rm PMNS}^\dagger$. For tiny neutrino masses, $x_i=m_{\nu_i}^2/M_W^2\ll 1$,
\bea
F_W(x)\to\frac{5}{6}-\frac{x}{4}+{\cal O}(x^2),
\eea
and we recover a well known result \cite{Cheng:1985bj} bounded by neutrino oscillation experiments:
\bea
{\cal B}(\muegamma)_{\rm SM}=\frac{3\alpha}{32\pi}
\left|\sum_i\bU^{ei}\bU^{\mu i*}\ x_i\right|^2 \lsim 10^{-54}.
\eea

\subsubsection*{Diagrams exchanging $Z_H$}
 
Taking now $M_1=M_{Z_H}$ and $M_2=m_{\ell_H^i}$, with the same $y_i$, we get:
\bea
\mbox{I:}&&
F_M^\gamma|_{Z_H}=-\ii F_E^\gamma|_{Z_H} = \frac{\alpha_W}{16\pi}m_\mu\sum_i V_{H\ell}^{ie*}V_{H\ell}^{i\mu} \ \left[C_0+3C_1+\frac{3}{2}C_{11}\right],
\\
\mbox{III:}&&
F_M^\gamma|_{Z_H}=-\ii F_E^\gamma|_{Z_H} = -\frac{\alpha_W}{32\pi}m_\mu\sum_i V_{H\ell}^{ie*}V_{H\ell}^{i\mu}\  y_i\ \left[C_1-\frac{3}{2}C_{11}\right],
\\
\mbox{Total:}&&
F_M^\gamma|_{Z_H}=-\ii F_E^\gamma|_{Z_H} = \frac{\alpha_W}{16\pi}\frac{m_\mu}{M_{W_H}^2}\sum_iV_{H\ell}^{ie*}V_{H\ell}^{i\mu}\ 
F_{\rm Z}(y_i),
\label{fmZ}
\eea
where
\bea
F_Z(x)&=&M_1^2\left[ C_0+3C_1+\frac{3}{2}C_{11}-\frac{x}{2}\left(C_1-\frac{3}{2}C_{11}\right)\right]
\nonumber \\
&=&
-\frac{1}{3}+\frac{2x+5x^2-x^3}{8(1-x)^3}+\frac{3x^2}{4(1-x)^4}\ln x,
\label{F_Z}
\eea
in agreement with \cite{Lavoura:2003xp,Blanke:2007db}.

\subsubsection*{Diagrams exchanging $A_H$}

This contribution can be obtained from that of the diagrams with a $Z_H$, 
replacing $Z_H$ by $A_H$. It is convenient to introduce the mass ratio
\bea
y'_i=ay_i,\quad
a=\frac{M_{W_H}^2}{M_{A_H}^2}=\frac{5c_W^2}{s_W^2}.
\eea
Then,
\bea
F_M^\gamma|_{A_H}=-\ii F_E^\gamma|_{A_H} &=& \frac{\alpha_W}{16\pi}\frac{m_\mu}{M_{A_H}^2}\frac{1}{25}\frac{s_W^2}{c_W^2}\sum_i V_{H\ell}^{ie*}V_{H\ell}^{i\mu}\ F_{\rm Z}(y_i')\nn\\
&=& 
\frac{\alpha_W}{16\pi}\frac{m_\mu}{M_{W_H}^2}\frac{1}{5}\sum_i V_{H\ell}^{ie*}V_{H\ell}^{i\mu}\ F_{\rm Z}(y_i').
\label{fmA}
\eea

\subsubsection*{Branching ratio}

Using $M_W^2/M_{W_H}^2=v^2/(4f^2)$ and $M_{W_H}=M_{Z_H}$, we finally obtain:
\bea
\label{muegammabr}
{\cal B}(\muegamma)&=&\frac{3\alpha}{2\pi}
\left|
\frac{v^2}{4f^2} \sum_i V_{H\ell}^{ie*}V_{H\ell}^{i\mu}\ 
\left(F_W(y_i)+F_Z(y_i)+\frac{1}{5}F_Z(ay_i)
\right)
\right|^2 ,
\eea
with $F_W$ and $F_Z$ given in Eqs.~(\ref{F_W}) and (\ref{F_Z}), respectively.

%%%%%%%%%%%%%%%%%%%%%%%%%%%%%%%%%%%%%%%%%%%%%%%%%%%%%%%%%%%%%%%%%%%%%%%%%%%%
\subsection{$\mueee$}

\begin{figure}
\begin{center}
\begin{tabular}{cccc}
\hspace{-8mm}
\includegraphics[scale=0.64]{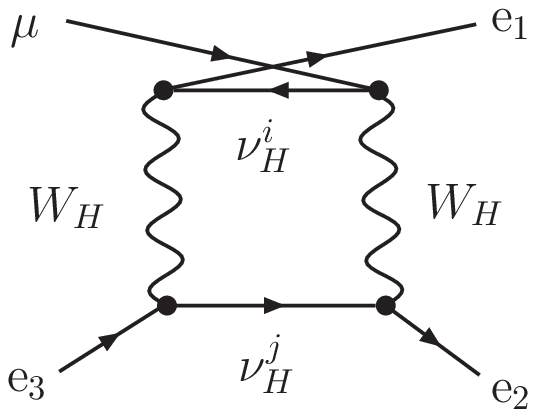}   & \hspace{-8mm}
\includegraphics[scale=0.64]{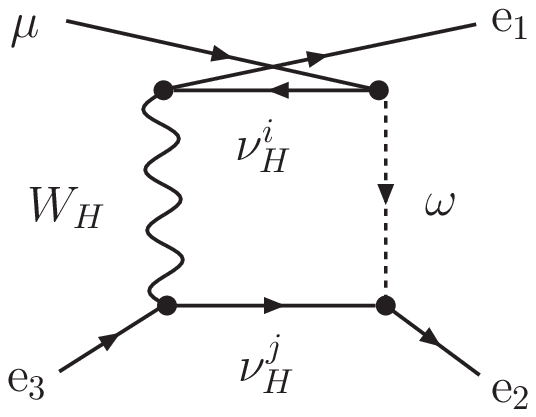}   & \hspace{-8mm}
\includegraphics[scale=0.64]{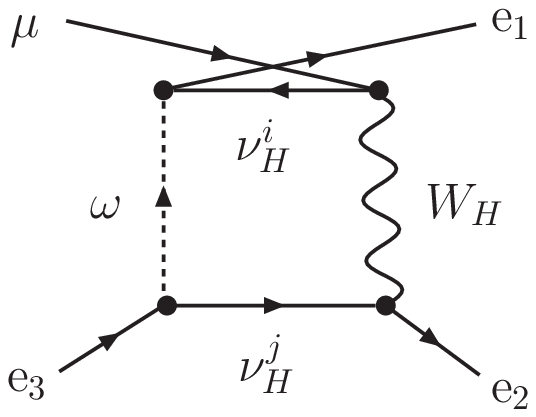}   & \hspace{-8mm}
\includegraphics[scale=0.64]{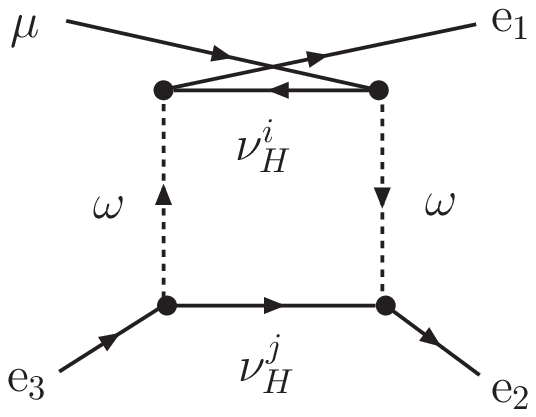}     \hspace{-8mm}
\\
\hspace{-8mm} A1a & \hspace{-8mm} A2a & \hspace{-8mm} A3a & \hspace{-8mm} A4a \\
\hspace{-8mm}
\includegraphics[scale=0.64]{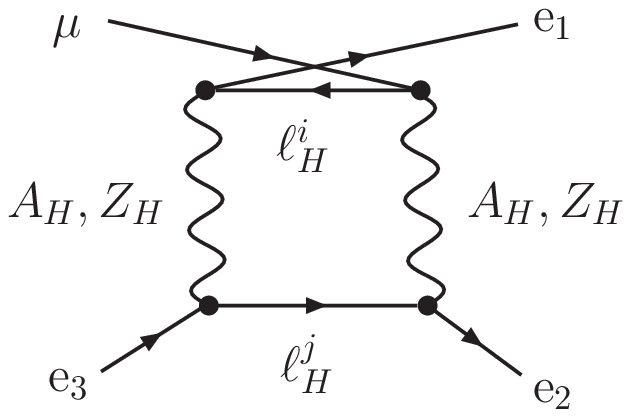} & \hspace{-8mm}
\includegraphics[scale=0.64]{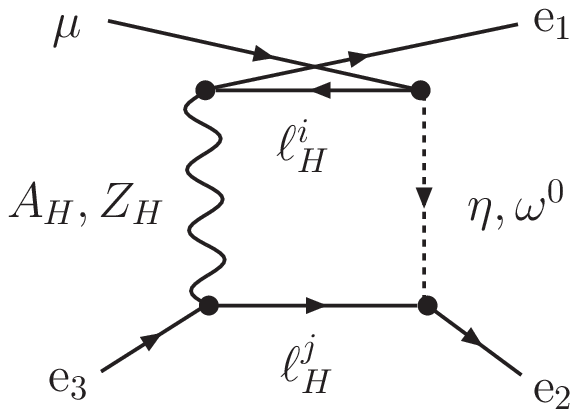} & \hspace{-8mm}
\includegraphics[scale=0.64]{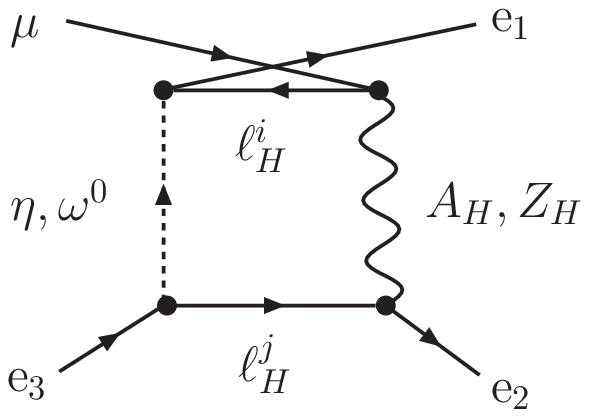} & \hspace{-8mm}
\includegraphics[scale=0.64]{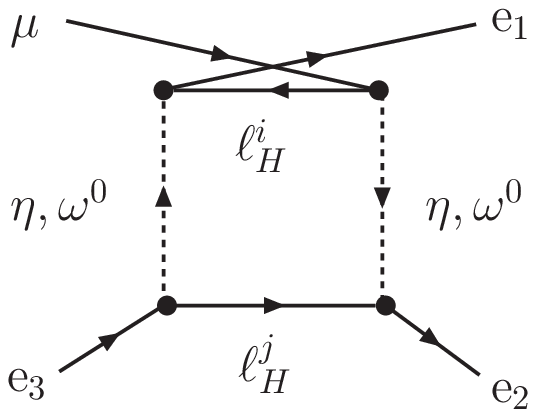}   \hspace{-8mm}
\\
\hspace{-8mm} A1b & \hspace{-8mm} A2b & \hspace{-8mm} A3b & \hspace{-8mm} A4b \\
\hspace{-8mm}
\includegraphics[scale=0.64]{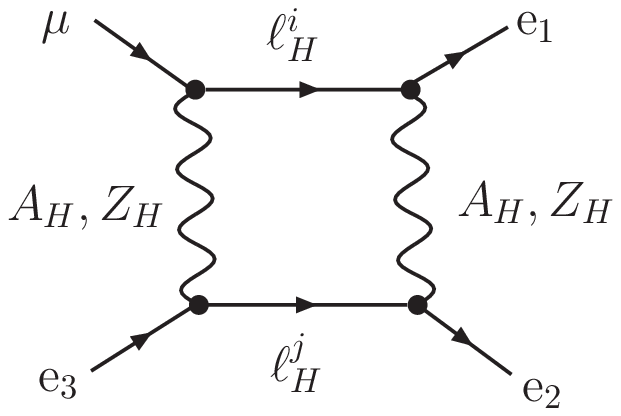} & \hspace{-8mm}
\includegraphics[scale=0.64]{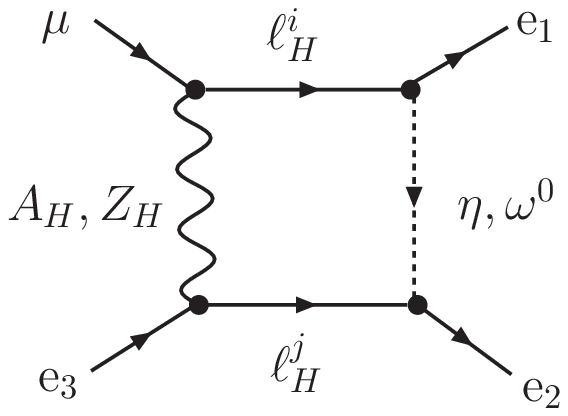} & \hspace{-8mm}
\includegraphics[scale=0.64]{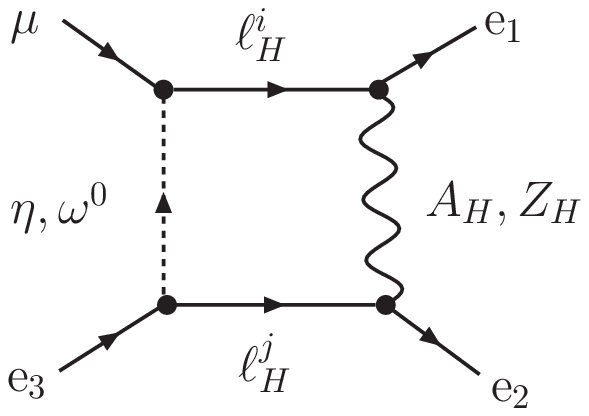} & \hspace{-8mm}
\includegraphics[scale=0.64]{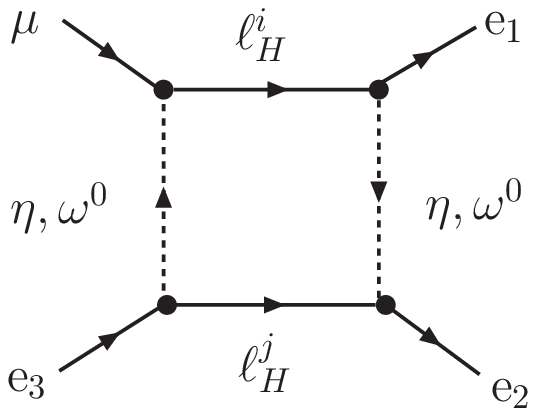}   \hspace{-8mm}
\\
\hspace{-8mm} B1 & \hspace{-8mm} B2 & \hspace{-8mm} B3 & \hspace{-8mm} B4 \\
\end{tabular}
\end{center}
\caption{Box diagrams for $\mu\to\e_1\e_2\bar\e_3$ in the LHT model. Crossed diagrams with $\e_1$ and $\e_2$ exchanged must be added.
\label{box-LH}}
\end{figure}

The self-energy diagrams do contribute to $F_{L,R}^V$ in this process, 
and must be included to calculate the penguin diagrams in $\mueee$ . 
On the other hand, 
apart from the box diagrams, only $\gamma$- and Z-penguin diagrams contribute 
to $\mueee$ in the LHT model. This is so because $A_H$ and $Z_H$ do not 
couple to two ordinary fermions, as required by T-parity conservation. 
(Higgs-penguins vanish in the limit of massless electrons, as do $F_R^V$ 
in this limit too.)
For the sake of brevity we present our results 
grouping together the $W_H$, $Z_H$ and $A_H$ contributions, but 
we distinguish among the $\gamma$- and $Z$-penguins in Fig.~\ref{Vff-LH} 
and the boxes in Fig.~\ref{box-LH}.

%%%%%%%%%%%%%%%  FIGURE %%%%%%%%%%%%%%%%%%%%%%%%%
%%%%%%%%%%%%%%%%%%%%%%%%%%%%%%%%%%%%%%%%%%%%%%%%%

%%%%%%%%%%%%%%%%%%%%%%%%%%%%%%%%%%%%%%%%%%%%%%%%%%%%%%%%%%%%%%%%%%%%%%%%%%%%
\subsubsection*{\boldmath{The $\gamma$-penguin}}

The form factors $F_M^\gamma$ and $F_E^\gamma$ have the same expressions (\ref{fmW},\ref{fmZ},\ref{fmA}) as for an on-shell photon, since terms of order $Q^2$ can be neglected. 
The contributions to $F_L^\gamma$, which are proportional to 
$Q^2 \sim m_\mu^2$ as expected, are detailed below.

For $M_1=M_{W_H}$, $M_2=m_{\nu_H^i}$ and $y_i=m_{Hi}^2/M_{W_H}^2$ as before,
the diagrams with $W_H$ yield
\bea
F_L^\gamma|_{W_H}&=&\frac{\alpha_W}{4\pi}\sum_i V_{H\ell}^{ie*}V_{H\ell}^{i\mu}\ 
G_W(y_i) \nonumber \\
&=&\frac{\alpha_W}{4\pi}\frac{Q^2}{M_{W_H}^2}\sum_i V_{H\ell}^{ie*}V_{H\ell}^{i\mu}\ 
G^{(1)}_W(y_i),
\eea
with
\bea
G_W(x)&=&\!\!-\frac{1}{2}+\overline B_1+6\overline C_{00} + x\left(\frac{1}{2}\overline B_1+\overline C_{00}-M_1^2\overline C_0\right)
-\left(2\overline C_1+\frac{1}{2}\overline C_{11}\right)Q^2
\label{gw}
\nn\\
&=&\Delta_\epsilon-\ln\frac{M_1^2}{\mu^2}+\frac{Q^2}{M_1^2}G^{(1)}_W(x)
+{\cal O}\left(\frac{Q^4}{M_1^4}\right),
\\
G^{(1)}_W(x)&=&
-\frac{5}{18}
+\frac{x(12+x-7x^2)}{24(1-x)^3}+\frac{x^2(12-10x+x^2)}{12(1-x)^4}\ln x.
\eea
Relations (\ref{li1}) and (\ref{li3}) have been used in (\ref{gw}).
Note that owing to the unitarity of the mixing matrix the $x$-independent terms in $G_W(x)$ drop out (including the ultraviolet divergence). The SM prediction is obtained by replacing $W_H$ by $W$, $\nu_H^i$ by $\nu_i$ and $V_{H\ell}$ by $V_{\rm PMNS}^\dagger$.

For $M_1=M_{Z_H}(=M_{W_H})$, $M_2=m_{\ell_H^i}\simeq m_{\nu_H^i}$ 
and the same $y_i$, the contribution of diagrams with $Z_H$ is
\bea
F_L^\gamma|_{Z_H}&=&\frac{\alpha_W}{4\pi}\sum_i V_{H\ell}^{ie*}V_{H\ell}^{i\mu}\ 
G_Z(y_i)\\
&=&\frac{\alpha_W}{4\pi}\frac{Q^2}{M_{W_H}^2}\sum_i V_{H\ell}^{ie*}V_{H\ell}^{i\mu}\ 
G^{(1)}_Z(y_i),
\eea
with
\bea
G_Z(x)&=&\left(1+\frac{x}{2}\right)\left(-\frac{1}{4}+\frac{1}{2}\overline B_1+C_{00}-\frac{x}{2}M_1^2 C_0\right) %\nn\\&&
-\left(\frac{1}{2}C_0+C_1+\frac{1}{8}(2+x)C_{11}\right)Q^2
\nn\\
&=&\frac{Q^2}{M_1^2}G^{(1)}_Z(x)+{\cal O}\left(\frac{Q^4}{M_1^4}\right),
\label{gz}
\\
G^{(1)}_Z(x)&=&
\frac{1}{36}
%+\frac{x(18-29x+10x^2+x^3)-2(4-16x+9x^2)\ln x}{48(1-x)^4}
+\frac{x(18-11x-x^2)}{48(1-x)^3}-\frac{4-16x+9x^2}{24(1-x)^4}\ln x.
\eea
The relation (\ref{li3}) has been used in Eq. (\ref{gz}).

Finally, the contribution of diagrams with $A_H$ is
obtained from that of diagrams with $Z_H$ 
replacing $Z_H$ by $A_H$, and $y_i$ by $y'_i = 5c_W^2 y_i / s_W^2$:
\bea
F_L^\gamma|_{A_H} = \frac{\alpha_W}{4\pi}
\frac{Q^2}{M_{W_H}^2}\frac{1}{5}\sum_i V_{H\ell}^{ie*}V_{H\ell}^{i\mu}\ G_{\rm Z}^{(1)}(y_i') .
\eea

%%%%%%%%%%%%%%%%%%%%%%%%%%%%%%%%%%%%%%%%%%%%%%%%%%%%%%%%%%%%%%%%%%%%%%%%%%%%
\subsubsection*{The Z-penguin}

The Z dipole form factors $F_{M,E}^Z$ (which are chirality flipping 
and hence proportional to the muon mass) can be neglected as compared 
to $F_L^Z$. This is in contrast with the $\gamma$-penguin, for which 
$Q F_{M,E}^\gamma (\sim Q F_{M,E}^Z) 
\sim Q^2/M_{W_H}^2\lsim m_\mu^2/M_{W_H}^2\sim F_L^\gamma$, to be 
compared with $F_L^Z\sim 1$. 
This justifies to neglect $F_{M,E}^Z$ in the $Z$-penguin (\ref{zpenguin}).

Taking $M_1=M_{W_H}$, $M_2=m_{Hi}$ and $y_i=m_{Hi}^2/M_{W_H}^2$, 
and using the unitarity of $V_{H\ell}$ we obtain:
\begin{align}
F_L^Z|_{W_H}&=&\frac{\alpha_W}{8\pi}\frac{1}{s_Wc_W}
\sum_i V_{H\ell}^{ie*}V_{H\ell}^{i\mu}
\bigg\{&
-2c_W^2\left(-\frac{1}{2}+\overline B_1+6\overline C_{00}-y_iM_{W_H}^2\overline C_0\right)
\nn\\
&&&-y_ic_W^2\left(\overline B_1+2\overline C_{00}\right)
\nn\\
&&&+2\left(1+\frac{y_i}{2}\right)\left(-\frac{1}{4}+\frac{1}{2}\overline B_1+C_{00}-\frac{y_i}{2}M_{W_H}^2C_0\right)
\nn\\
&&&+\frac{v^2}{f^2}\frac{y_i}{16}\left[1+4(\overline C_{00}-C_{00}+M_{W_H}^2\left( C_0-2\overline C_0\right))\right]\bigg\}
\nonumber \\
&=&\frac{\alpha_W}{8\pi}\frac{1}{s_Wc_W}
\sum_i V_{H\ell}^{ie*}V_{H\ell}^{i\mu}
\bigg\{&-2c_W^2\left(\Delta_\epsilon-\ln\frac{M_{W_H}^2}{\mu^2}\right)
+\frac{v^2}{f^2}\frac{y_i}{8}\ H_W(y_i) \bigg\}
\nn\\
&=&\frac{\alpha_W}{8\pi}\frac{1}{s_Wc_W}
\sum_i V_{H\ell}^{ie*}V_{H\ell}^{i\mu} &\frac{v^2}{f^2}\frac{y_i}{8}\
H_W(y_i)\ ,
\end{align}
with
\bea
H_W(x)=\frac{6-x}{1-x}+\frac{2+3x}{(1-x)^2}\ln x , 
\eea
and
\bea
F_L^Z|_{Z_H}&=&\frac{\alpha_W}{8\pi}\frac{1}{s_Wc_W}
\sum_i V_{H\ell}^{ie*}V_{H\ell}^{i\mu}
(1-2c_W^2)
\left(-\frac{1}{4}+\frac{1}{2}\overline B_1+C_{00}-\frac{y_i}{2}M_{W_H}^2 C_0\right)
\nn\\
&&\times\left\{
\left(1+\frac{y_i}{2}\right)
-\frac{v^2}{f^2}\left[\frac{y_i}{4}+\left(\frac{c_W}{s_W}y_i-\frac{2s_W}{5c_W}\right)x_H\right]\right\}=0,
\\
F_L^Z|_{A_H}&=&\frac{\alpha_W}{8\pi}\frac{1}{s_Wc_W}
\sum_i V_{H\ell}^{ie*}V_{H\ell}^{i\mu}
(1-2c_W^2)
\left(-\frac{1}{4}+\frac{1}{2}\overline B_1+C_{00}-\frac{y_i}{2}M_{A_H}^2 C_0\right)
\nn\\
&&\times\frac{1}{25}\frac{s_W^2}{c_W^2}\left\{
\left(1+\frac{y_i'}{2}\right)
-\frac{v^2}{f^2}\left[\frac{5}{4}y_i'+\left(\frac{s_W}{c_W}y_i'+10\frac{c_W}{s_W}\right)x_H\right]\right\}=0 . 
\eea
Here $x_H$ is a constant defining the mixing between the heavy neutral 
gauge bosons and function of the gauge couplings (see Eq. (\ref{XH})).
We observe that the only contribution to the Z-penguins comes from the diagrams with $W_H$, 
and it is proportional to $v^2/f^2$. The potentially dangerous ultraviolet divergences proportional to $y_i$ have cancelled thanks to the proper $v^2/f^2$ corrections to the 
$\omega^\pm W_H^\mp Z$ and $Z\bar\nu_{HR}^i\nu_{HR}^i$ couplings. The corrections to the latter were not included in \cite{Blanke:2007db}.

For completeness, we give the prediction for the Z-penguin in the SM with light 
massive neutrinos. Although in the LHT the heavy leptons are vector-like and 
the Z boson couples to both chiralities, the final form of the vertex 
is the same. This is more easily seen in the unitary gauge, where the 
heavy modes contribution is only given by diagram I in Fig.~\ref{Vff-LH} 
and is proportional to the $v^2/f^2$ correction to the 
$Z\bar\nu_{HR}^i\nu_{HR}^i$ coupling 
(see \cite{Blanke:2006eb} for further discussion).
Taking $M_1=M_W$, $M_2=m_{\nu_i}$ and $x_i=m_{\nu_i}^2/M_W^2$, 
and using the unitarity of $V_{\rm PMNS}$ we obtain:
\begin{align}
F_L^Z|_W&=&\frac{\alpha_W}{8\pi}\frac{1}{s_Wc_W}
\sum_i V_{\rm PMNS}^{ei}V_{\rm PMNS}^{\mu i*}
\bigg\{&
-2c_W^2\left(-\frac{1}{2}+\overline B_1+6\overline C_{00}-x_iM_W^2\overline C_0\right)
\nn\\
&&&+\frac{x_i}{2}(1-2c_W^2)\left(\overline B_1+2\overline C_{00}\right)
\nn\\
&&&-\frac{1}{2}+\overline B_1+2C_{00}-\frac{x_i}{2}M_W^2(4\overline C_0+x_i C_0)\bigg\}
\nn\\
&=&\frac{\alpha_W}{16\pi}\frac{1}{s_Wc_W}
\sum_i V_{\rm PMNS}^{ei}V_{\rm PMNS}^{\mu i*}
\bigg\{&
-4c_W^2\left(\Delta_\epsilon-\ln\frac{M_W^2}{\mu^2}\right)-1 
\nn\\
&&&
+2(\overline B_1+2C_{00})-x_iM_W^2(4\overline C_0+x_i C_0)\bigg\}
\nn\\
&=&\frac{\alpha_W}{16\pi}\frac{1}{s_Wc_W}
\sum_i V_{\rm PMNS}^{ei}V_{\rm PMNS}^{\mu i*}\
&x_iH_W(x_i),
\end{align}
which is, of course, finite and in agreement with Ref.~\cite{Illana:2000ic} for $Q^2=0$.

%%%%%%%%%%%%%%%%%%%%%%%%%%%%%%%%%%%%%%%%%%%%%%%%%%%%%%%%%%%%%%%%%%%%%%%%%%%%
\subsubsection*{Box diagrams}

There are eight different classes of box diagrams grouped in types A and B 
in the LHT model (Fig.~\ref{box-LH}). 
In the limit of zero external momenta 
(all internal masses are much larger than the muon mass) 
all of them have the same form, being proportional to a scalar integral 
over the internal momentum $q$.  
Indeed, omitting the corresponding denominator 
$(q^2-m_{Hi}^2)^2(q^2-M^2_{G_H})^2$, with $G=W,Z$ or $A$, 
{\allowdisplaybreaks
\bea
{\rm A1:} &&
\bra{p_1}\gamma^\mu P_L(-\slash{q}+m_{Hi})\gamma^\nu P_L\ket{p}
\bra{p_2}\gamma_\nu P_L(-\slash{q}+m_{Hi})\gamma_\mu P_L\ket{p_3} \nn\\
&&= \frac{q^2}{4}
\bra{p_1}\gamma^\mu\gamma^\alpha\gamma^\nu P_L\ket{p}
\bra{p_2}\gamma_\nu\gamma_\alpha\gamma_\mu P_L\ket{p_3} \nn\\
&&= q^2
\bra{p_1}\gamma^\mu P_L\ket{p}\bra{p_2}\gamma_\mu P_L\ket{p_3},
\\
{\rm A2:} &&
-\bra{p_1}\gamma^\mu P_L(-\slash{q}+m_{Hi}) P_L\ket{p}
\bra{p_2}P_R(-\slash{q}+m_{Hi})\gamma_\mu P_L\ket{p_3} \nn\\
&&=
-m_{Hi} m_{Hj}\bra{p_1}\gamma^\mu P_L\ket{p}\bra{p_2}\gamma_\mu P_L\ket{p_3}, 
\\
{\rm A3:} &&
-\bra{p_1}P_R(-\slash{q}+m_{Hi})\gamma^\mu P_L\ket{p}
\bra{p_2}\gamma_\mu P_L(-\slash{q}+m_{Hi})P_L\ket{p_3} \nn\\
&&=
-m_{Hi} m_{Hj}\bra{p_1}\gamma^\mu P_L\ket{p}\bra{p_2}\gamma_\mu P_L\ket{p_3},
\\
{\rm A4:} &&
\bra{p_1} P_R(-\slash{q}+m_{Hi})P_L\ket{p}
\bra{p_2} P_R(-\slash{q}+m_{Hi})P_L\ket{p_3} \nn\\
&&= \frac{q^2}{4}
\bra{p_1}\gamma^\mu P_L\ket{p}\bra{p_2}\gamma_\mu P_L\ket{p_3}, 
\eea
\bea
{\rm B1:} &&
\bra{p_1}\gamma^\mu P_L(\slash{q}+m_{Hi})\gamma^\nu P_L\ket{p}
\bra{p_2}\gamma_\nu P_L(-\slash{q}+m_{Hi})\gamma_\mu P_L\ket{p_3} \nn\\
&&= -\frac{q^2}{4}
\bra{p_1}\gamma^\mu\gamma^\alpha\gamma^\nu P_L\ket{p}
\bra{p_2}\gamma_\mu\gamma_\alpha\gamma_\nu P_L\ket{p_3} \nn\\
&&= -4q^2
\bra{p_1}\gamma^\mu P_L\ket{p}\bra{p_2}\gamma_\mu P_L\ket{p_3},
\\
{\rm B2:} &&
-\bra{p_1}\gamma^\mu P_L(\slash{q}+m_{Hi}) P_L\ket{p}
\bra{p_2}P_R(-\slash{q}+m_{Hi})\gamma_\mu P_L\ket{p_3} \nn\\
&&=
-m_{Hi} m_{Hj}\bra{p_1}\gamma^\mu P_L\ket{p}\bra{p_2}\gamma_\mu P_L\ket{p_3},
\\
{\rm B3:} &&
-\bra{p_1}P_R(\slash{q}+m_{Hi})\gamma^\mu P_L\ket{p}
\bra{p_2}\gamma_\mu P_L(-\slash{q}+m_{Hi})P_L\ket{p_3} \nn\\
&&=
-m_{Hi} m_{Hj}\bra{p_1}\gamma^\mu P_L\ket{p}\bra{p_2}\gamma_\mu P_L\ket{p_3}, 
\\
{\rm B4:} &&
\bra{p_1} P_R(\slash{q}+m_{Hi})P_L\ket{p}
\bra{p_2} P_R(-\slash{q}+m_{Hi})P_L\ket{p_3} \nn\\
&&=-\frac{q^2}{4}
\bra{p_1}\gamma^\mu P_L\ket{p}\bra{p_2}\gamma_\mu P_L\ket{p_3}.
\eea}
Thus, all box form factors except $B_1^L$ vanish (see Eq. (\ref{MB})). 
Using the Fierz identity
\bea
\bra{1}\gamma^\mu P_L\ket{2}\bra{3}\gamma_\mu P_L\ket{4}
=-\bra{3}\gamma^\mu P_L\ket{2}\bra{1}\gamma_\mu P_L\ket{4}
\eea
and including all the factors, we obtain the generic expressions 
for the contributions from diagrams of types A and B 
(see Fig.~\ref{box-LH} and Appendix~\ref{appfr} for definitions), 
{\allowdisplaybreaks
\bea
{\rm A:}\quad B_1^L&=&
2\frac{\alpha}{4\pi}\sum_{ij}\bigg[\left(g_{L1}^{ie*}g_{L2}^{i\mu}g_{L1}^{je}g_{L2}^{je*}
+\frac{1}{4}c_{L1}^{ie*}c_{L2}^{i\mu}c_{L1}^{je}c_{L2}^{je*}\right)\widetilde D_0(M_1^2,M_2^2,m_{Hi}^2,m_{Hj}^2)
\nn\\
&&-\left(g_{L1}^{ie*}c_{L2}^{i\mu}g_{L1}^{je}c_{L2}^{je*}
+c_{L1}^{ie*}g_{L2}^{i\mu}c_{L1}^{je}g_{L2}^{je*}\right)
m_{Hi}m_{Hj}
D_0(M_1^2,M_2^2,m_{Hi}^2,m_{Hj}^2)\bigg],\quad\quad
\\
{\rm B:}\quad B_1^L&=&
2\frac{\alpha}{4\pi}\sum_{ij}\bigg[-\left(4g_{L2}^{ie*}g_{L1}^{i\mu}g_{L1}^{je}g_{L2}^{je*}
+\frac{1}{4}c_{L2}^{ie*}c_{L1}^{i\mu}c_{L1}^{je}c_{L2}^{je*}\right)\widetilde D_0(M_1^2,M_2^2,m_{Hi}^2,m_{Hj}^2)
\nn\\
&&-\left(g_{L2}^{ie*}c_{L1}^{i\mu}c_{L1}^{je}g_{L2}^{je*}
+c_{L2}^{ie*}g_{L1}^{i\mu}g_{L1}^{je}c_{L2}^{je*}\right)
m_{Hi}m_{Hj}
D_0(M_1^2,M_2^2,m_{Hi}^2,m_{Hj}^2)\bigg].\quad\quad
\eea}
\noindent
Finally, replacing the vertex coefficients given in Appendix~\ref{appfr} we 
derive the contributions of the heavy gauge bosons and the corresponding 
would-be-Goldstone bosons:
{\allowdisplaybreaks
\bea
B_1^L(W_H,W_H)&=&
\frac{\alpha}{2\pi}\frac{1}{4s_W^4}\frac{1}{M_W^2}\frac{v^2}{4f^2}\sum_{ij}\chi_{ij}\left[\left(1+\frac{1}{4}y_iy_j\right)\widetilde d_0(y_i,y_j)
-2y_iy_j d_0(y_i,y_j)\right],\quad\quad
\\
B_1^L(Z_H,Z_H)&=&
\frac{\alpha}{2\pi}\frac{1}{16s_W^4}\frac{1}{M_W^2}\frac{v^2}{4f^2}\sum_{ij}\chi_{ij}\left[-3\widetilde d_0(y_i,y_j)\right],
\\
B_1^L(A_H,A_H)&=&
\frac{\alpha}{2\pi}\frac{1}{16s_W^4}\frac{1}{25a}\frac{1}{M_W^2}\frac{v^2}{4f^2}\sum_{ij}\chi_{ij}\left[-3\widetilde d_0(y_i',y_j')\right],
\\
B_1^L(Z_H,A_H)&=&
\frac{\alpha}{2\pi}\frac{1}{16s_W^4}\frac{2}{5}\frac{1}{M_W^2}\frac{v^2}{4f^2}\sum_{ij}\chi_{ij}\left[-3\widetilde d_0(a,y_i',y_j')\right],
\eea}
with
\bea
\chi_{ij}&=&V_{H\ell}^{ie*}V_{H\ell}^{i\mu}|V_{H\ell}^{je}|^2.
\eea
The SM contribution from the exchange of the light neutrinos is 
again similar to that from $W_H$, but performing the corresponding 
replacements.

\subsubsection*{Branching ratio}

Collecting everything, the non-vanishing contributions to the vertex and 
box form factors in Eq.~(\ref{mueee}) from $\gamma$-penguins, 
Z-penguins and box diagrams in the LHT can be written 
\bea
A_1^L=\frac{F_L^\gamma}{Q^2}&=&
\frac{\alpha_W}{4\pi}\frac{1}{M_W^2}\frac{v^2}{4f^2}\sum_i V_{H\ell}^{ie*}V_{H\ell}^{i\mu}
\left[G^{(1)}_W(y_i)+G^{(1)}_Z(y_i)+\frac{1}{5}G^{(1)}_Z(ay_i)
\right],
\\
A_2^R=\frac{2 F_M^\gamma}{m_\mu}&=&
\frac{\alpha_W}{8\pi}\frac{1}{M_W^2}\frac{v^2}{4f^2}\sum_i V_{H\ell}^{ie*}V_{H\ell}^{i\mu}
\left[F_W(y_i)+F_Z(y_i)+\frac{1}{5}F_Z(ay_i)
\right],
\\
F_{LL}=-\frac{F_L^ZZ_L^e}{M_Z^2}&=&
\frac{\alpha_W}{8\pi}\frac{1-2s_W^2}{2s_W^2}\frac{1}{M_W^2}\frac{v^2}{f^2}\sum_i 
V_{H\ell}^{ie*}V_{H\ell}^{i\mu} 
\frac{y_i}{8}\ H_W(y_i),
\\
F_{LR}=-\frac{F_L^ZZ_R^e}{M_Z^2}&=&
-\frac{\alpha_W}{8\pi}\frac{1}{M_W^2}\frac{v^2}{f^2}\sum_i 
V_{H\ell}^{ie*}V_{H\ell}^{i\mu} \frac{y_i}{8}\
H_W(y_i),
\\
B_1^L&=&
\frac{\alpha_W}{8\pi}\frac{1}{s^2_W}\frac{1}{M_W^2}\frac{v^2}{4f^2}\sum_{ij}\chi_{ij}
\bigg[\left(1+\frac{1}{4}y_iy_j\right)\tilde d_0(y_i,y_j)-2y_iy_jd_0(y_i,y_j)
\nn\\
&&\hspace{1cm}-\frac{3}{4}\tilde d_0(y_i,y_j)-\frac{3}{100a}\tilde d_0(ay_i,ay_j)
-\frac{3}{10}\tilde d_0(a,ay_i,ay_j)\bigg].
\eea
The branching ratio reads 
\begin{align}
{\cal B}(\mu&\to \e\e\bar\e)=12 s_W^4 M_W^4\Bigg\{
|A_1^L|^2
-2(A_1^LA_2^{R*}+{\rm h.c.})
+|A_2^R|^2\left(\frac{16}{3}\ln\frac{m_\mu}{m_e}-\frac{22}{3}\right)
\nn\\ 
&+\frac{1}{6}|B_1^L|^2+\frac{1}{3}(A_1^LB_1^{L*}
+{\rm h.c.})
-\frac{2}{3}(A_2^RB_1^{L*}+{\rm h.c.}) +\frac{1}{3}\left(2|F_{LL}|^2+|F_{LR}|^2\right)\nn\\
&+\frac{1}{3}\left(B_1^LF_{LL}^*
+2A_1^LF_{LL}^*+A_1^LF_{LR}^*-4A_2^RF_{LL}^*-2A_2^RF_{LR}^*+{\rm h.c.}\right)
\bigg\}.
\end{align}

%%%%%%%%%%%%%%%%%%%%%%%%%%%%%%%%%%%%%%%%%%%%%%%%%%%%%%%%%%%%%%%%%%%%%%%%%%%%%%%%%%%%%%%%%%%%%%%%%%%%%%%%%%%%
\section{Numerical results}
\label{limits}

In order to study the bounds on the new parameters imposed by 
the experimental limits on $\muegamma$ and $\mueee$, it is convenient to restrict 
ourselves to the case of two generations. 
Hence, we are left with four parameters: the LH order parameter $f$, 
the masses of the two heavy lepton doublets in (\ref{lHmasses}) $m_{Hi}$ ($i=1,2$), 
and the angle $\theta$ defining the $2\times 2$ 
mixing matrix between the heavy and the SM charged leptons
\bea
V_{H\ell}
=\left(\ba{cc} V_{H\ell}^{1e} & V_{H\ell}^{1\mu} \\ 
V_{H\ell}^{2e} & V_{H\ell}^{2\mu} \ea\right)
=\left(\ba{rr} \cos\theta & \sin\theta \\ -\sin\theta & \cos\theta \ea\right).  
\eea 
(In the contributions we consider the $e$ and $\mu$ phases, as well as the 
heavy lepton doublet phases, can be safely redefined.)
We shall replace $m_{H1},\ m_{H2}$ by $\delta$ and $\tilde y$, 
however, to present our results. 
The former, which is proportional to the heavy lepton mass difference, 
describes together with $\theta$ the alignment between heavy and SM charged leptons,
\bea
\delta=\frac{m_{H2}^2-m_{H1}^2}{m_{H1}m_{H2}}. 
\eea
Whereas the latter, which sets the heavy lepton scale, is relevant for discussing 
decoupling,  
\bea
\tilde y=\sqrt{y_1y_2},\quad
y_i=\frac{m_{Hi}^2}{M_{W_H}^2},\quad
i=1,2. 
\eea
Note that both $M_{W_H}$ and $m_{Hi}$ are proportional to $f$ (see Appendix \ref{apppf}). 
The penguin contributions then take the form
\bea
\label{mue}
\sum_{i=1}^2 V_{H\ell}^{ie*} V_{H\ell}^{i\mu}\ F(y_i)
=\frac{\sin 2\theta}{2}\left[F(y_1)-F(y_2)\right], 
\eea
where $F$ stands for a generic function; 
and the box contributions 
\begin{align}
\label{mueeed}
\sum_{i,j=1}^2 V_{H\ell}^{ie*}V_{H\ell}^{i\mu}
|V_{H\ell}^{je}|^2\ F(y_i,y_j)=\frac{\sin 2\theta}{2}\big\{&\cos^2\theta\left[F(y_1,y_1)-F(y_2,y_1)\right]\nn\\
+&\sin^2\theta\left[F(y_1,y_2)-F(y_2,y_2)\right]\big\}.
\end{align}
Thus, the LFV amplitudes vanish for vanishing mixing, $\theta = 0$, or heavy mass 
splitting, $\delta = 0$.

%%%%%%%%%%%%%%%  FIGURE %%%%%%%%%%%%%%%%%%%%%%%%%
\begin{figure}
\begin{center}
\includegraphics[scale=1]{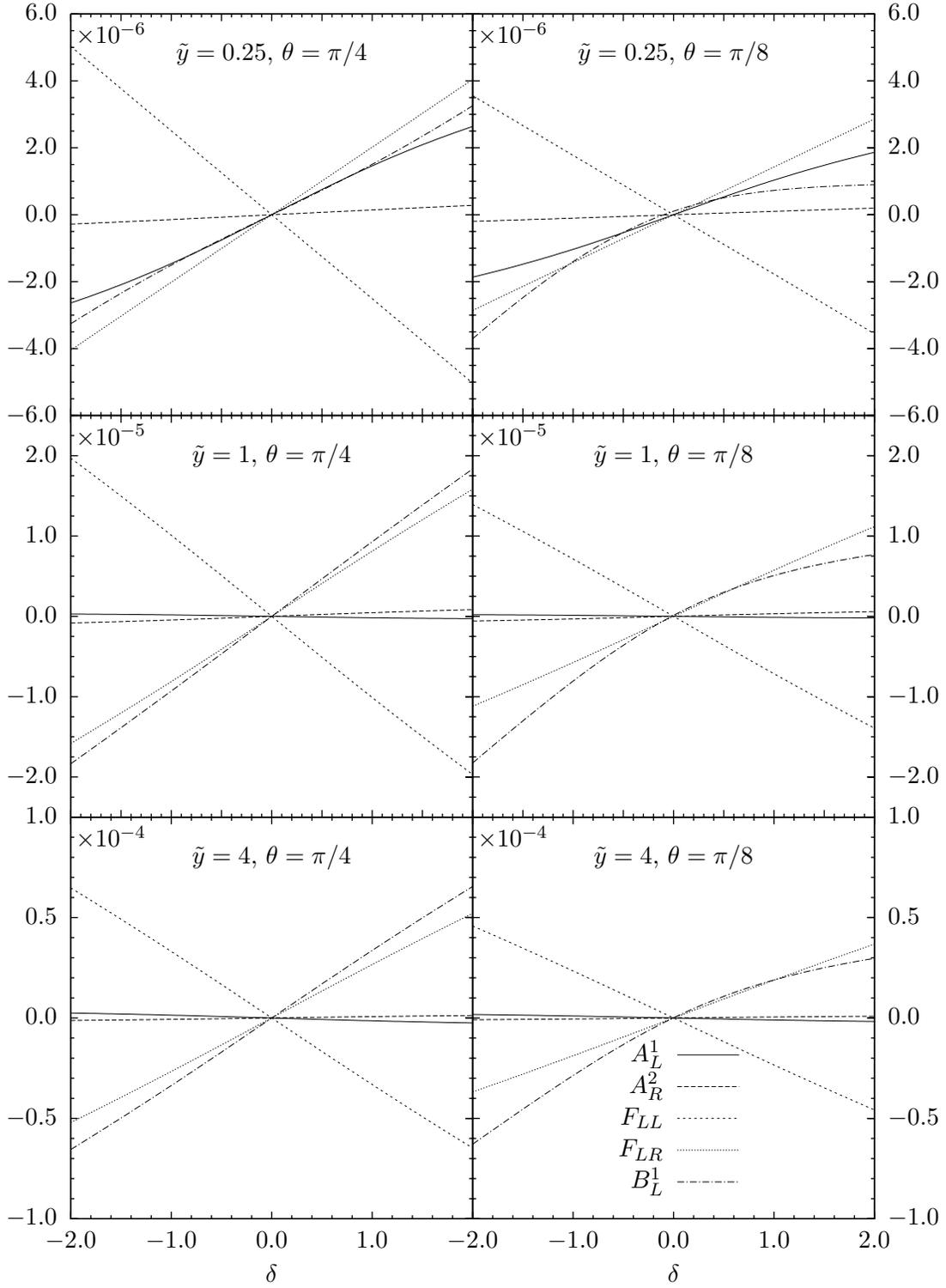}
\end{center}
\caption{
Form factors multiplied by $M_W^2$, for $f=1$ TeV.
\label{fig:ff}}
\end{figure}
%%%%%%%%%%%%%%%%%%%%%%%%%%%%%%%%%%%%%%%%%%%%%%%%%

\begin{figure}
\begin{center}
\includegraphics[scale=1]{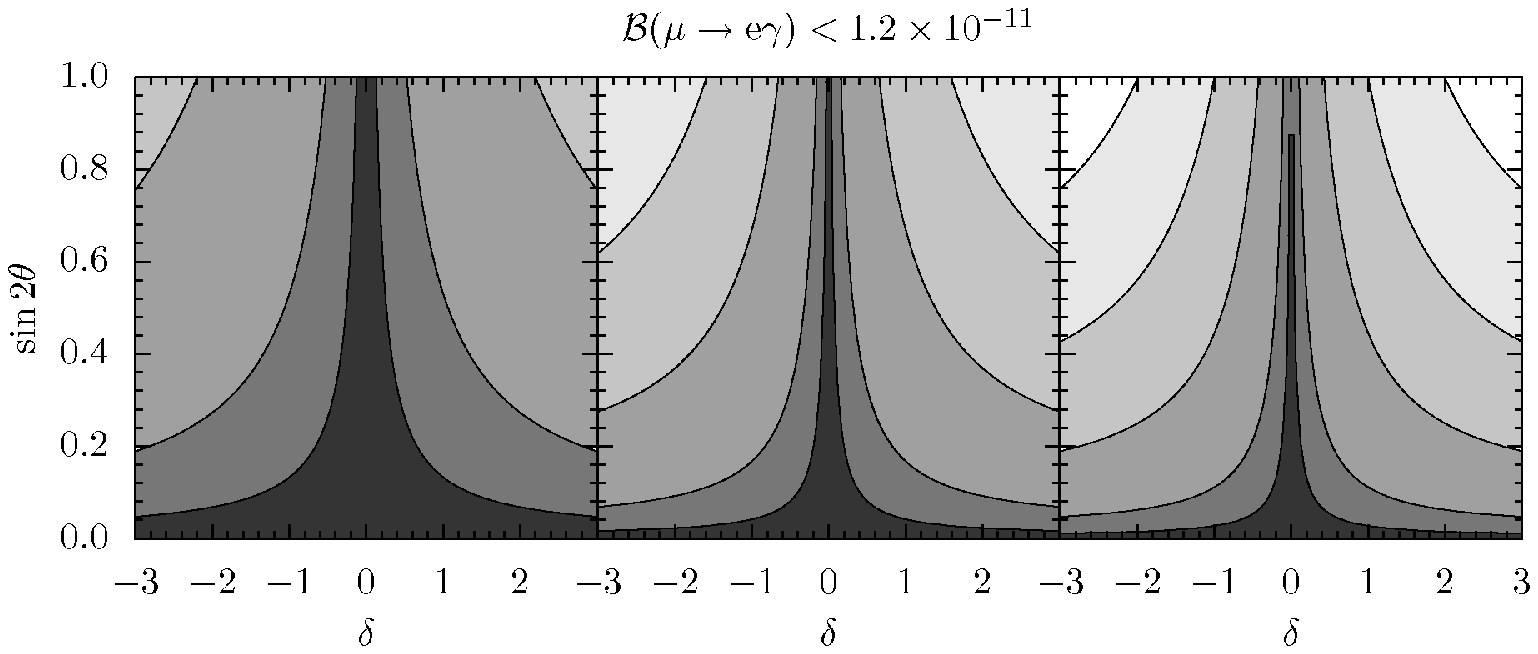}
\end{center}
\caption{
Contours of ${\cal B}(\muegamma)=1.2\times10^{-11}$ in the $(\sin2\theta,\delta)$ plane for $\tilde y=0.25,1,4$ (left, center, right) and $f=0.5,1,2,3,4$ TeV (from bottom up). 
\label{fig:muegamma}}
\end{figure}

\begin{figure}
\begin{center}
\includegraphics[scale=1]{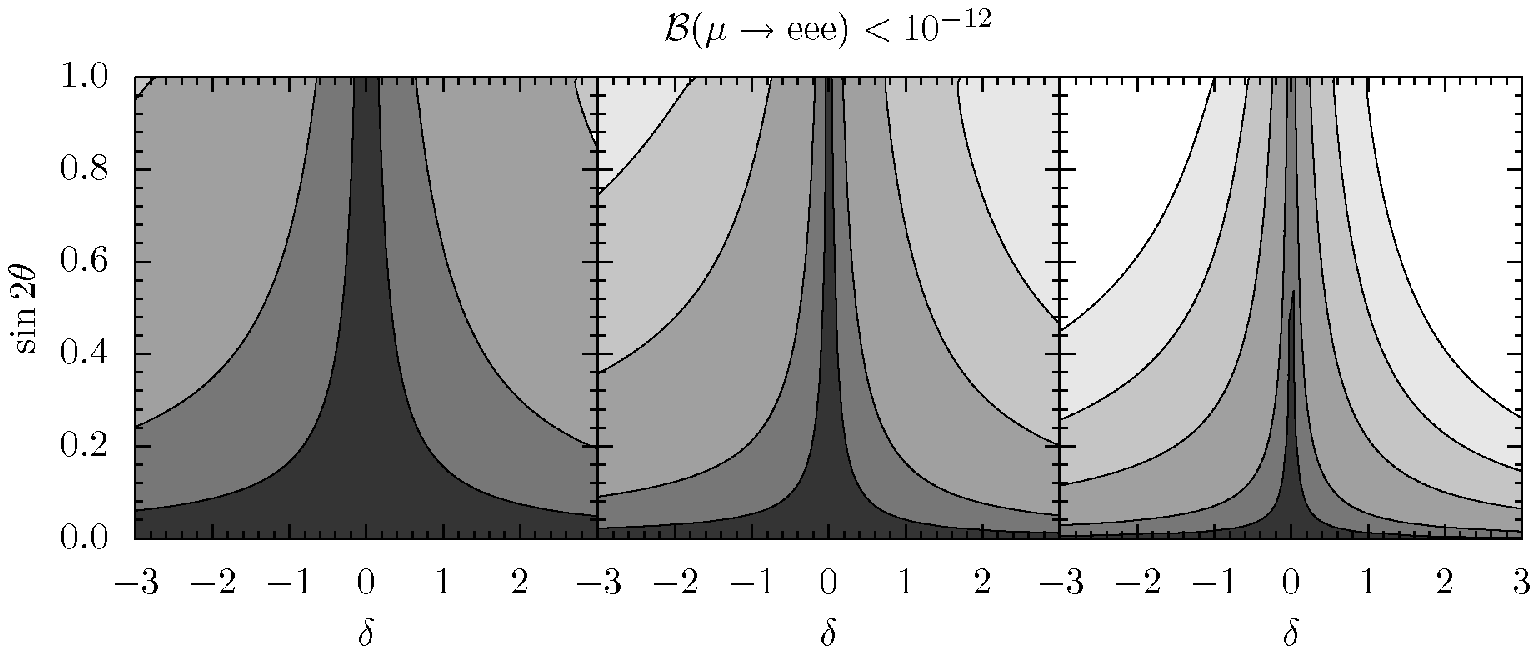}
\end{center}
\caption{
Contours of ${\cal B}(\mueee)=10^{-12}$ in the $(\sin2\theta,\delta)$ plane for $\tilde y=0.25,1,4$ (left, center, right) and $f=0.5,1,2,3,4$ TeV (from bottom up).
\label{fig:mueee}}
\end{figure}

We plot for illustration in Fig.~\ref{fig:ff} the form factors
for the $\muegamma$ and $\mueee$ decay amplitudes calculated in 
the previous section as a function of $\delta$ 
for several $\theta$ and $\tilde y$ values and $f=1$ TeV. 
They grow with $\tilde y$ and scale like $f^{-2}$. 
In contrast with the MSSM case \cite{Hisano:1995cp,Arganda:2005ji}, 
box contributions to $\mueee$ are of the same order than penguins, 
in particular for $\tilde y\gsim 1$, which explains the different
behaviour of the decay rates with the sign of $\delta$ for non-maximal 
flavour mixing.
The dependence on the new parameters is more clearly seen 
in Figs.~\ref{fig:muegamma} and \ref{fig:mueee}. They show the
present exclusion contours in the $(\sin 2\theta,\delta)$ plane 
implied by the present limits on  
${\cal B}(\muegamma)<1.2\times 10^{-11}$ \cite{Brooks:1999pu}
and ${\cal B}(\mueee)<10^{-12}$ \cite{Bellgardt:1987du}, respectively,
and for three values of $\tilde y$, 0.25, 1, 4.
The regions above each line of constant $f$ are excluded.
As it can be observed, mixing angle and mass splitting are correlated, 
because the alignment between 
the Yukawa couplings of the heavy and the SM charged leptons 
goes to zero with any of them. 
Present limits on LFV muon decays imply that $\theta$ 
or $\delta \lsim 0.1$ for $\tilde y = 1$ and $f=1$ TeV. 
If no LFV signal is seen by the MEG experiment at PSI,
the limits are expected to improve by two orders of
magnitude \cite{RittMori}
and the corresponding exclusion contours would be
those in  Figs.~\ref{fig:muegamma} and \ref{fig:mueee}
replacing $f$ by $\sqrt{10}f$.

%%%%%%%%%%%%%%%  FIGURE %%%%%%%%%%%%%%%%%%%%%%%%%
%%%%%%%%%%%%%%%%%%%%%%%%%%%%%%%%%%%%%%%%%%%%%%%%%

%%%%%%%%%%%%%%%  FIGURE %%%%%%%%%%%%%%%%%%%%%%%%%
%%%%%%%%%%%%%%%%%%%%%%%%%%%%%%%%%%%%%%%%%%%%%%%%%

\begin{figure}
\begin{center}
\begin{tabular}{cc}
\includegraphics[scale=1]{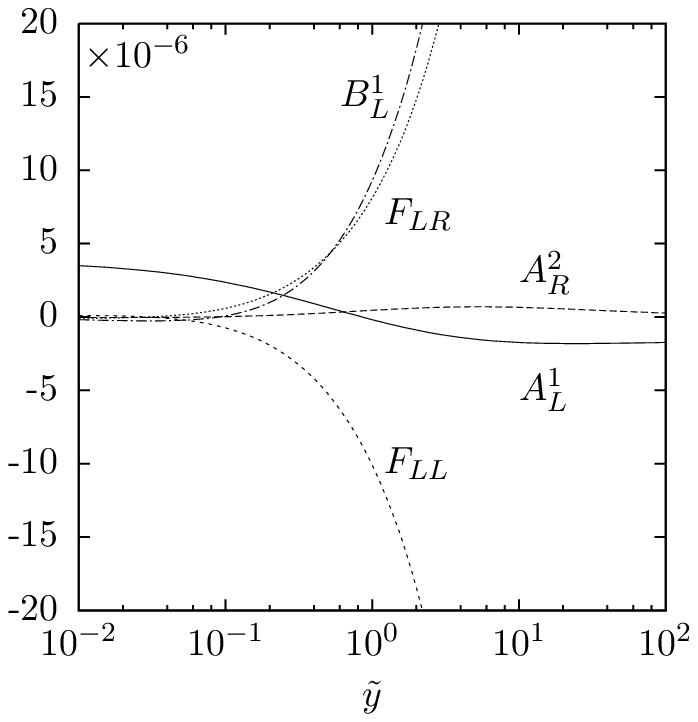} &\hskip4mm
\includegraphics[scale=1]{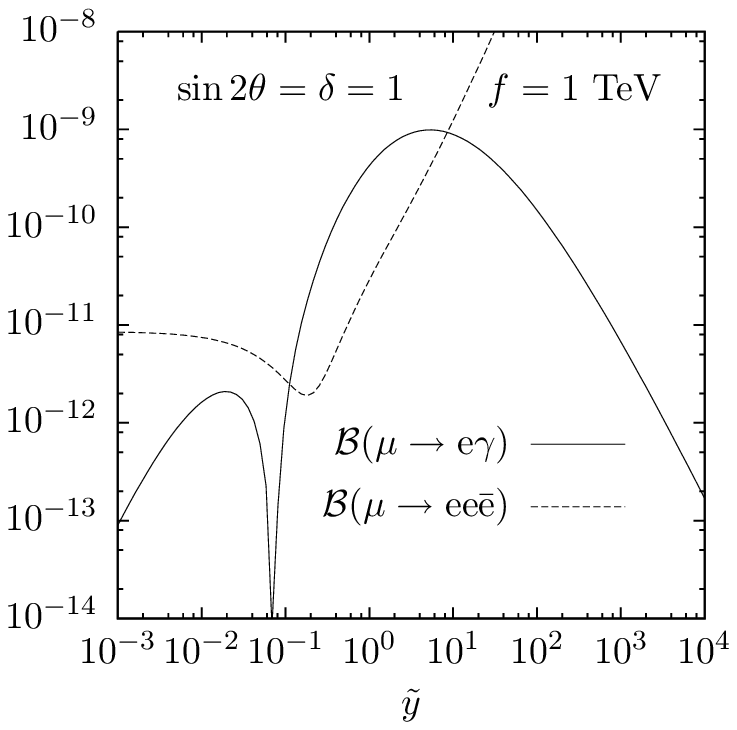}
\end{tabular}
\end{center}
\caption{
Form factors multiplied by $M_W^2$ (left) and branching ratios (right) as a function of
$\tilde y$ for $\sin2\theta=\delta=1$ and $f=1$~TeV. The latter must be compared with 
present experimental limits on ${\cal B}(\muegamma)<1.2\times 10^{-11}$ 
and ${\cal B}(\mueee)<10^{-12}$.
\label{fig:tildey}}
\end{figure}

As already emphasized, 
the LFV branching ratios scale like $f^{-4}$. However, 
the $\tilde y$ dependence deserves more discussion. 
In Fig.~\ref{fig:tildey} we plot the variation of the form 
factors and of the branching ratios 
with $\tilde y$ for maximal mixing, $\sin 2\theta = 1$, 
and $\delta = 1$. Two comments are in order. The 
non-observation of these LFV processes already sets 
non-trivial limits on the LHT parameters because 
the central region $\tilde y \sim 1$ is already excluded 
for natural values of the other parameters. 
More interestingly, 
${\cal B}(\mueee)$ goes like $\tilde y^2$ for 
very large $\tilde y$. 
This is so because $\tilde y$ is quadratic in the heavy 
Yukawa coupling $\kappa$, which goes to infinity 
with the heavy lepton masses for fixed $f$. 
This behaviour is similar to the leading EWPD 
dependence on the top quark mass \cite{Sirlin:1980nh}, which scales 
with $m_t^2$ in the region of physical interest, 
allowing a determination of the top mass from a global 
fit \cite{Amsler:2008zz}. 
Just like in the top quark case, the dependence is moderate
when the particles within multiplets become degenerate (the 
symmetry is recovered). Generic limits from all these 
figures are tabulated in the summary below. 

%%%%%%%%%%%%%%%  FIGURE %%%%%%%%%%%%%%%%%%%%%%%%%
%%%%%%%%%%%%%%%%%%%%%%%%%%%%%%%%%%%%%%%%%%%%%%%%%

Let us, finally, comment on the general case with three families. 
Similarly to the two family case, we have to align the new 
contributions to the electron and to the muon at the 10\% level. 
However, now this alignment is not easily related to the usual 
parameterization of the mixing in terms of two mass splittings, 
three mixing angles and one phase 
(as in the two family case we can safely redefine the $e$ and $\mu$, 
as well as the heavy lepton doublet phases, and then use the same 
parameterization as for the CKM matrix \cite{Amsler:2008zz}). 
In order to estimate the fine tuning required by 
$\muegamma$, for instance, we rather introduce the ratio 
(see Eq. (\ref{mue}))
\bea
\label{muealign}
\frac{\left|\displaystyle\sum_{i=1}^3 V_{H\ell}^{ie*} V_{H\ell}^{i\mu}\ F(y_i)\right|^2}
{\left(\displaystyle\sum_{i=1}^3 |V_{H\ell}^{ie}|^2\ |F(y_i)|\right)
\left(\displaystyle\sum_{i=1}^3 |V_{H\ell}^{i\mu}|^2\ |F(y_i)|\right)},    
\eea
which approximately scales like $\sin^2 2\theta\delta^2$ for two families. 
This is $\lsim 10^{-2}$ for almost 
all the three family parameter space when ${\cal B}(\muegamma)$ is below 
the present experimental limit, and at most $\sim 10^{-4}$ if the limit 
improves by two orders of magnitude. There is a special region in 
parameter space, however, where the ratio (\ref{muealign}) can be larger 
even though ${\cal B}(\muegamma)$ is well below the experimental limit. 
This is around $y_i \sim 0.3$, where the total amplitude 
$F(y_i) \sim F_W(y_i)+F_Z(y_i)+\frac{1}{5}F_Z(ay_i)$ 
in Eq. (\ref{muegammabr}) is negligible 
(see right panel of Fig.~\ref{fig:tildey}).  
But this region is excluded by the present limit on ${\cal B}(\mueee)$.
Thus, the new contributions to the electron and to the muon 
must be aligned at the 10\% level, being the square of this precision 
the largest value of the ratio in Eq. (\ref{muealign}). 
Analogously, we can define the corresponding ratio using the 
amplitudes for $\mueee$ in Eqs.~(\ref{mue},\ref{mueeed}), obtaining 
similar results. 
The numerical analysis presented here is at some extent complementary 
to the study in Ref.~\cite{Blanke:2007db}, where the correlation 
between different observables, in particular between 
${\cal B}(\muegamma)$ and ${\cal B}(\mueee)$, is explicitly shown.

%%%%%%%%%%%%%%%%%%%%%%%%%%%%%%%%%%%%%%%%%%%%%%%%%%%%%%%%%%%%%%%%%%%%%%%%%%%%
\section{Conclusions}
\label{conclusions}

LH models provide a natural explanation of the little 
hierarchy between the EW scale and the scale where we expect the NP to be, and 
which is to be explored by the LHC. 
However, these models where the Higgs is a pseudo-Goldstone boson of an approximate global 
symmetry in general suffer some tension in accommodating the many new 
particles required near the TeV scale without upsetting the EWPD constraints. 
This is ameliorated by further extending the model to include a discrete
symmetry, the T parity, under which all observed particles, including the Higgs boson, 
are even and hopefully all the new ones are odd.
All these models, as any {\em universal} NP near the TeV scale, must also 
guarantee that the new particles do not mediate too large FCNC processes. 
We have recalculated the new contributions to the LFV processes 
$\muegamma$ and $\mueee$ in the LHT model, the most economical of such proposals.
The full Lagrangian has been introduced and all pieces of the calculation, 
in particular the Z-penguin and box diagrams contributing to $\mueee$, 
have been considered in detail. We have found that the former are ultraviolet 
finite when all the Goldstone boson interactions to the order considered are included. 
Whereas we recover previous results for $\muegamma$ 
\cite{Choudhury:2006sq,Blanke:2007db}.

The present limits on the rates of LFV processes
translate into bounds on the LHT parameters. Tables~\ref{tab1} and
\ref{tab2} show the bounds 
imposed in the two family case by $\muegamma$ and $\mueee$, 
respectively, on the heavy lepton mass splitting $\delta$ 
for a maximal mixing angle, $\sin 2\theta = 1$, and several 
values of the LH scale $f$ and the ratio $\tilde y$ 
related to the common heavy lepton mass. 
The main conclusion is that the new parameters must be tuned 
to 10\% for a {\em natural} value $f\sim1$~TeV. Obviously, raising $f$ 
quickly reduces the decay rates, which scale as $f^{-4}$. 
The results are also sensitive to the parameter 
$\tilde y$, but the dependence is mild for moderate values 
(see Fig.~\ref{fig:tildey}), when rates scale roughly like 
$\sin^2 2\theta\ \delta^2$. The non-observation of LFV effects may 
be also the result of a conspiracy among the new parameters 
being all slightly above or below their expected natural values, 
of order one. 
%Analogous fine tuning is required in the general case with 
%three families but on the generalised mixing definition.
Analogous fine tuning on the alignment of light and heavy leptons
is required in the general case with three families.

%%%%%%%%%%%%%%%  TABLE %%%%%%%%%%%%%%%%%%%%%%%%%
\begin{table}
\begin{center}
\begin{tabular}{|c|ccc|}
\hline
\multicolumn{4}{|c|}{${\cal B}(\muegamma)<1.2\times10^{-11}$\quad ($\sin2\theta=1$)} \\
\hline
$f$ [TeV] & $\tilde y=0.25$ & $\tilde y=1$ & $\tilde y=4$ \\
\hline
0.5 & $|\delta|<0.13$ & $|\delta|<0.040$ & $|\delta|<0.027$ \\
1.0 & $|\delta|<0.52$ & $|\delta|<0.16$ & $|\delta|<0.11$ \\
2.0 & $|\delta|<2.2$ & $|\delta|<0.66$ & $|\delta|<0.43$ \\
4.0 & $|\delta|<14$ & $|\delta|<3.5$ & $|\delta|<2.0$ \\
\hline
\end{tabular}
\end{center}
\caption{Bounds on the splitting $\delta$ from the present experimental limit on
 ${\cal B}(\muegamma)$ for several scales $f$ and ratios $\tilde y$, taking
 $\sin2\theta=1$.\label{tab1}}
\end{table}
%%%%%%%%%%%%%%%%%%%%%%%%%%%%%%%%%%%%%%%%%%%%%%%%

%%%%%%%%%%%%%%%  TABLE %%%%%%%%%%%%%%%%%%%%%%%%%
\begin{table}[htb]
\begin{center}
\begin{tabular}{|c|ccc|}
\hline
\multicolumn{4}{|c|}{${\cal B}(\mueee)<10^{-12}$\quad ($\sin2\theta=1$)} \\
\hline
$f$ [TeV] & $\tilde y=0.25$ & $\tilde y=1$ & $\tilde y=4$ \\
\hline
0.5 & $|\delta|<0.16$ & $|\delta|<0.045$ & $|\delta|<0.015$ \\
1.0 & $|\delta|<0.64$ & $|\delta|<0.18$ & $|\delta|<0.061$ \\
2.0 & $|\delta|<2.7$ & $|\delta|<0.72$ & $|\delta|<0.24$ \\
4.0 & $|\delta|<13$ & $|\delta|<3.3$ & $|\delta|<0.98$ \\
\hline
\end{tabular}
\end{center}
\caption{Bounds on the splitting $\delta$ from the present experimental limits on
 ${\cal B}(\mueee)$ for several scales $f$ and ratios $\tilde y$, taking
 $\sin2\theta=1$.\label{tab2}}
\end{table}
%%%%%%%%%%%%%%%%%%%%%%%%%%%%%%%%%%%%%%%%%%%%%%%%

%%%%%%%%%%%%%%%  TABLE %%%%%%%%%%%%%%%%%%%%%%%%%
\begin{table}[htb]
\begin{center}
\begin{tabular}{|r|rl|rl|}
\hline
& \multicolumn{2}{c|}{${\cal B}(\muegamma)<1.2\times10^{-11}$ ($10^{-13}$)} & 
  \multicolumn{2}{c|}{${\cal B}(\mueee)<10^{-12}$ ($10^{-14}$)}\\
\hline
$f/\mbox{TeV}>$ & 2.5  & (8.1)   & 2.3  & (7.4) \\
$\sin2\theta<$  & 0.16 & (0.015) & 0.16 & (0.016) \\
$|\delta|<$     & 0.16 & (0.015) & 0.18 & (0.018) \\
$\tilde y<$     & 0.16 & (*)     & *    & (*) \\
\hline
\end{tabular}
\end{center}
\caption{Bounds from current (future) experiments on individual LHT parameters, assuming the others fixed to the {\it natural} values $f=1$ TeV, $\sin2\theta=\delta=\tilde y=1$. 
An asterisk means that the quoted limit on the branching ratio excludes 
any $\tilde y$ value for the assumed values of the other parameters. 
\label{tab3}}
\end{table}
%%%%%%%%%%%%%%%%%%%%%%%%%%%%%%%%%%%%%%%%%%%%%%%%

In Table \ref{tab3} we give both the present and future bounds 
if the current limits on $\muegamma$ and $\mueee$ are improved by 
two orders of magnitude \cite{RittMori}. An asterisk indicates that the assumed 
values are excluded for any possible $\tilde y$. A non-empty 
region for $\tilde y$ is recovered increasing $f$ or decreasing 
$\sin2\theta$ and/or $\delta$ (see Figs.~\ref{fig:muegamma} and \ref{fig:mueee}).
Finally, we must note that the limits on the corresponding tau decays 
$\tau\to\mu\gamma,\e\gamma$ and $\tau\to\mu\mu\bar\mu,\e\e\bar{\e}$ 
are weaker \cite{taudecays}, typically $< 10^{-8}-10^{-7}$.
Then, they do not further restrict the order parameter $f$ 
for natural values of the other heavy lepton parameters, 
but could eventually constrain the corresponding mixing angles and 
heavy lepton masses, which are in principle independent of the parameters 
otherwise involved in the muon to electron processes. 
However, present limits give no significative bound on the parameters related 
to the third lepton family.

%%%%%%%%%%%%%%%%%%%%%%%%%%%%%%%%%%%%%%%%%%%%%%%%%%%%%%%%%%%%%%%%%%%%%%%%%%%%
\noindent
{\it Note added:}

During the completion of this manuscript several related papers 
were released.
The one-loop contributions in the LHT to the $tbW$ vertex in 
Ref.~\cite{Penunuri:2008pb} and to $Z\ell\ell'$ in 
Ref.~\cite{Yue:2008rh} have been calculated. 
In neither case has the order $v^2/f^2$ correction to the SM weak boson 
coupling to heavy right-handed fermions been included. 
More recently a new analysis of $B$ decays 
in the this model has been carried out in 
Ref.~\cite{Altmannshofer:2008dz}, yielding an ultraviolet 
finite result when this correction was taken into account 
following Ref.~\cite{Goto:2008fj} and in agreement with our findings.

%%%%%%%%%%%%%%%%%%%%%%%%%%%%%%%%%%%%%%%%%%%%%%%%%%%%%%%%%%%%%%%%%%%%%%%%%%%%

%%%%%%%%%%%%%%%%%%%%%%%%%%%%%%%%%%%%%%%%%%%%%%%%%%%%%%%%%%%%%%%%%%%%%%%%%%%%
\subsection*{Acknowledgements}

Very useful discussions with Jorge de Blas, Mar{\'\i}a Jos\'e Herrero, 
Manuel Masip, Manuel P\'erez-Victoria, 
Antonio Pich, Cecilia Tarantino and Jos\'e Wudka are gratefully acknowledged.
This work has been supported by MICINN project FPA2006-05294 and
Junta de Andaluc{\'\i}a projects FQM 101, FQM 437 and FQM03048.
The work of M.D.J. has been supported by a MICINN FPU fellowship.

%%%%%%%%%%%%%%%%%%%%%%%%%%%%%%%%%%%%%%%%%%%%%%%%%%%%%%%%%%%%%%%%%%%%%%%%%%%%
\appendix
\renewcommand{\arraystretch}{2}

\section{Physical fields\label{apppf}}

After the electroweak symmetry breaking (EWSB) the SM gauge boson 
mass eigenstates, which are the T-even, write 
\bea
W^\pm=\frac{1}{\sqrt{2}}(W^1\mp\ii W^2),\quad
\left(\ba{c} Z \\ A \ea\right)=\left(\ba{cc} c_W & s_W \\ -s_W & c_W \ea\right)
\left(\ba{c} W^3 \\ B \ea\right) ,
\eea
with
\bea
W^a=\frac{W_1^a+W_2^a}{\sqrt{2}},\quad
B=\frac{B_1+B_2}{\sqrt{2}} ;
\eea
whereas the T-odd combinations expanding the heavy sector read to order $v^2/f^2$ are
\bea
W_H^\pm=\frac{1}{\sqrt{2}}(W_H^1\mp\ii W_H^2),\quad
\left(\ba{c} Z_H \\ A_H \ea\right)=\left(\ba{cc} 1 & -x_H\dis\frac{v^2}{f^2} \\ x_H\dis\frac{v^2}{f^2} & 
1 \ea\right)\left(\ba{c} W_H^3 \\ B_H \ea\right),
\eea
with
\bea
W_H^a=\frac{W_1^a-W_2^a}{\sqrt{2}},\quad
B_H=\frac{B_1-B_2}{\sqrt{2}},\quad
x_H=\frac{5gg'}{4(5g^2-g'^2)}.
\label{XH}
\eea
Their masses to order $v^2/f^2$ are
\begin{align}
&M_W=\frac{gv}{2}\left(1-\frac{v^2}{12f^2}\right),\quad
M_Z=M_W/c_W,\quad
e=gs_W=g'c_W,\quad v\simeq246\mbox{ GeV}, &
\nn\\
&M_{W_H}=M_{Z_H}=gf\left(1-\frac{v^2}{8f^2}\right),\quad 
M_{A_H}=\frac{g'f}{\sqrt{5}}\left(1-\frac{5v^2}{8f^2}\right).&
\end{align}

The scalar fields must be also rotated into the physical fields
\cite{Hubisz:2005tx}:
{\allowdisplaybreaks
\bea
\pi^0 &\to& \pi^0\left(1+\frac{v^2}{12f^2}\right) , \\
\pi^\pm &\to& \pi^\pm\left(1+\frac{v^2}{12f^2}\right) , \\
h &\to& h , \\
\Phi^0 &\to& \Phi^0\left(1+\frac{v^2}{12f^2}\right) , \\
\Phi^P &\to& \Phi^P + \left(\sqrt{10}\eta-\sqrt{2}\omega^0+
\Phi^P\right)\frac{v^2}{12f^2} , \\
\Phi^\pm &\to& \Phi^\pm\left(1+\frac{v^2}{24f^2}\right)
\pm\ii\omega^\pm\frac{v^2}{12f^2} , \\
\Phi^{++} &\to& \Phi^{++} , \\
\eta &\to& \eta + \frac{5g'\eta-4\sqrt{5}[g'(\omega^0+\sqrt{2}\Phi^P)
-6gx_H\omega^0]}{24g'}\frac{v^2}{f^2} , \\
\omega^0 &\to& \omega^0 + \frac{5g(\omega^0+4\sqrt{2}\Phi^P)
-4\sqrt{5}\eta(5g+6g'x_H)}{120g}\frac{v^2}{f^2} , \\
\omega^\pm &\to& \omega^\pm\left(1+\frac{v^2}{24f^2}\right)
\pm\ii\Phi^\pm\frac{v^2}{f^2} .
\eea}

For each SM left-handed lepton doublet there is an extra vector-like 
doublet, 
\bea
l_{iL}=\left(\ba{c} \nu_{iL} \\ \ell_{iL}\ea\right),\quad i=1,2, \quad
l_{HR}=\left(\ba{c} \nu_{HR} \\ \ell_{HR}\ea\right) .
\eea
Then the left-handed mass eigenstates are
\bea
l_L=\frac{l_{1L}-l_{2L}}{\sqrt{2}},\quad
l_{HL}=\frac{l_{1L}+l_{2L}}{\sqrt{2}},\quad
l = \nu,\ell,
\eea
where we omit the flavour index. 
$\nu_L, \ell_L$ are the SM (T-even) left-handed leptons, 
whereas $\nu_{H L},\ell_{H L}$ ($\nu_{H R},\ell_{H R}$) are T-odd 
left (right) handed leptons with masses of ${\cal O}(f)$.
The SM right-handed fermions are assumed to be singlets under the 
non-abelian symmetries. 
Heavy leptons receive their masses from the Yukawa term proportional to 
$\kappa$ (\ref{mirror}), which is in general a non-diagonal matrix in flavour space 
that induces flavour mixing in the T-odd sector. 
The misalignment between the mass matrices of the T-even (SM) and 
T-odd (heavy) sectors is a source of intergeneration mixing 
(see Section~\ref{flavour}). 
The diagonalisation of the $\kappa$ matrix (see Eq. (\ref{Hmixing})) yield 
the heavy lepton masses
\bea
m_{\ell_H^i}=\sqrt{2}\kappa_{ii} f\equiv m_{Hi},\quad
m_{\nu_H^i}=m_{Hi}\left(1-\frac{v^2}{8f^2}\right).
\label{lHmasses}
\eea

\section{Feynman rules\label{appfr}}

We present below just the Feynman rules which are necessary for the calculation 
of the LFV processes discussed in this work. They are given in terms of generic 
couplings for the following general vertices involving scalars (S), fermions (F) 
and/or gauge bosons (V):
\bea
\mbox{[V$_\mu$FF]} &=& \ii e\gamma^\mu(g_LP_L+g_RP_R),
\\
\mbox{[SFF]} &=& \ii e(c_LP_L+c_RP_R),
\\
\mbox{[SV$_\mu$V$_\nu$]} &=& \ii eKg^{\mu\nu},
\\
\mbox{[V$_\mu$S$(p_1)$S$(p_2)$]} &=& \ii eG(p_1-p_2)^\mu,
\\
\mbox{[V$_\mu(p_1)$V$_\nu(p_2)$V$_\rho(p_3)$]} &=& \ii eJ\left[g^{\mu\nu}(p_2-p_1)^\rho + g^{\nu\rho}(p_3-p_2)^\mu + g^{\mu\rho}(p_1-p_3)^\nu\right] ,
\eea
where all momenta are assumed incoming. 
The conjugate vertices are obtained replacing:
\bea
g_{L,R}\leftrightarrow g_{L,R}^*,\quad
c_{L,R}\leftrightarrow c_{R,L}^*,\quad
K\leftrightarrow K^*,\quad
G\leftrightarrow G^*,\quad
J\leftrightarrow J^*.
\eea

\subsection{SM with massive neutrinos}

For comparison we first give the rules for the SM with light massive neutrinos. 
The sign conventions for the covariant derivatives are those in 
Ref.~\cite{Denner:1991kt}.

\noindent
\begin{tabular}{|c|c|c|c|c|}
\hline
VFF & $\gamma\bar f^if^j$ & $Z\bar f^if^j$  & $W^+\bar \nu^i\ell^j$ & $W^-\bar \ell^j\nu^i$
\\
\hline\hline
$g_L$ & $-Q_f\delta_{ij}$ & $Z_L^f\delta_{ij}$ & $\dis\frac{1}{\sqrt{2}s_W}\bU^{ji*}$ & $\dis\frac{1}{\sqrt{2}s_W}\bU^{ji}$
\\
%\hline
$g_R$ &$-Q_f\delta_{ij}$ & $Z_R^f\delta_{ij}$ & 0 & 0
\\
\hline 
\end{tabular}

\noindent
where  $Z_{L,R}^f=(v_f\pm a_f)/2s_Wc_W$ with $v_f=T_3^{f_L}-2Q_fs_W^2$ and $a_f=T_3^{f_L}$.

\noindent
\begin{tabular}{|c|c|}
\hline
SFF & $\phi^+\bar \nu^i\ell^j$ 
\\
\hline\hline
$c_L$ & 
$+\dis\frac{1}{\sqrt{2}s_W}\frac{m_{\nu^i}}{M_W}\bU^{ji*}$ 
\\
%\hline
$c_R$ & 
$-\dis\frac{1}{\sqrt{2}s_W}\frac{m_{\ell^j}}{M_W}\bU^{ji*}$ 
\\
\hline 
\end{tabular}

\noindent
\begin{tabular}{|c|c|c|}
\hline
SVV & $\phi^\pm W^\mp \gamma$ & $\phi^\pm W^\mp Z$ 
\\
\hline\hline
$K$ & $-M_W$ & $-M_W s_W/c_W$ 
\\
\hline 
\end{tabular}

\noindent
\begin{tabular}{|c|c|c|}
\hline
VSS & $\gamma\phi^\pm\phi^\mp$ & $Z\phi^\pm \phi^\mp$ \\
\hline\hline
$G$ & $\mp1$ & $\pm\dis\frac{c_W^2-s_W^2}{2s_Wc_W}$ 
\\
\hline 
\end{tabular}

\noindent
\begin{tabular}{|c|c|c|}
\hline
VVV & $\gamma W^+W^-$ & $ZW^+W^-$ \\
\hline\hline
$J$ & $-1$ & $c_W/s_W$
\\
\hline 
\end{tabular}

\noindent
The fields $\phi^\pm$ are the would-be Goldstone bosons eaten by the gauge bosons fields
$W^\pm$ after the EWSB.

\subsection{LHT model}

The sign conventions are chosen to be compatible with those 
employed for the SM (which coincide with those in \cite{Blanke:2006eb} 
up to a sign in the definition of the abelian gauge couplings in the covariant 
derivative in Eq. (\ref{derivative})). In particular, these Feynman rules include 
the ${\cal O}(v^2/f^2)$ contribution to the $Z\bar \nu_H^i\nu_H^j$ vertex missed in the literature.  

\noindent
\begin{tabular}{|c|c|c|c|}
\hline
VFF & $\gamma\bar f_H^if_H^j$ & $Z\bar \nu_H^i\nu_H^j$ & $Z\bar \ell_H^i\ell_H^j$ 
\\
\hline\hline
$g_L$ & $-Q_f\delta_{ij}$ &
$\dis\frac{1}{2s_Wc_W}\delta_{ij}$ & $\dis\frac{1}{2s_Wc_W}(-1+2s_W^2)\delta_{ij}$ 
\\
%\hline
$g_R$ &$-Q_f\delta_{ij}$ &
$\dis\frac{1}{2s_Wc_W}\left(1-\frac{v^2}{4f^2}\right)\delta_{ij}$ & $\dis\frac{1}{2s_Wc_W}(-1+2s_W^2)\delta_{ij}$ 
\\
\hline 
\end{tabular}

\noindent
\begin{tabular}{|c|c|c|c|}
\hline
VFF & $A_H\bar \ell_H^i\ell^j$ & $Z_H\bar \ell_H^i\ell^j$ &
$W_H^+\bar \nu_H^i\ell^j$ 
\\
\hline\hline
$g_L$ & 
$\left(\dis\frac{1}{10c_W}-\frac{x_H}{2s_W}\frac{v^2}{f^2}\right)
V_{H\ell}^{ij}$ & 
$-\left(\dis\frac{1}{2s_W}+\frac{x_H}{10c_W}\frac{v^2}{f^2}\right)
V_{H\ell}^{ij}$ & 
$\dis\frac{1}{\sqrt{2}s_W}V_{H\ell}^{ij}$ 
\\
%\hline
$g_R$ & 0 & 0 & 0 
\\
\hline 
\end{tabular}

\noindent
\begin{tabular}{|c|c|c|}
\hline
SFF & $\eta\bar \ell_H^i\ell^j$ & $\omega^0\bar \ell_H^i\ell^j$
\\
\hline\hline 
$c_L$ & 
$\dis\frac{\ii}{10c_W}\frac{m_{\ell_H^i}}{M_{A_H}}\left[1-\frac{v^2}{f^2}
\left(\frac{5}{4}\!+\!x_H\frac{s_W}{c_W}\right)\right]
V_{H\ell}^{ij}$ &
$\dis\frac{\ii}{2s_W}\frac{m_{\ell_H^i}}{M_{Z_H}}\left[1+\frac{v^2}{f^2}
\left(-\frac{1}{4}\!+\!x_H\frac{c_W}{s_W}\right)\right]
V_{H\ell}^{ij}$ 
\\
%\hline
$c_R$ & 
$-\dis\frac{\ii}{10c_W}\frac{m_{\ell^i}}{M_{A_H}}V_{H\ell}^{ij}$ &
$-\dis\frac{\ii}{2s_W}\frac{m_{\ell^i}}{M_{Z_H}}V_{H\ell}^{ij}$ 
\\
\hline 
\end{tabular}

\noindent
\begin{tabular}{|c|c|}
\hline
SFF & $\omega^+\bar \nu_H^i\ell^j$ 
\\
\hline\hline
$c_L$ & 
$-\dis\frac{\ii}{\sqrt{2}s_W}\frac{m_{\nu_H^i}}{M_{W_H}}
V_{H\ell}^{ij}$ 
\\
%\hline
$c_R$ & 
$\dis\frac{\ii}{\sqrt{2}s_W}\frac{m_{\ell^i}}{M_{W_H}}V_{H\ell}^{ij}$ 
\\
\hline 
\end{tabular}

\noindent
\begin{tabular}{|c|c|c|}
\hline
SVV & $\omega^\pm W_H^\mp \gamma$ & $\omega^\pm W_H^\mp Z$ 
\\
\hline\hline
$K$  & $\pm\ii M_{W_H}$ & $ \mp\ii M_{W_H} \dis\frac{c_W}{s_W}\left(1-\frac{v^2}{4f^2c^2_W}\right)$ 
\\
\hline 
\end{tabular}

\noindent
\begin{tabular}{|c|c|c|}
\hline
VSS & $\gamma\omega^\pm\omega^\mp$ & $Z\omega^\pm \omega^\mp$ \\
\hline\hline
$G$ & $\mp1$ & $\pm\dis\frac{c_W}{s_W}\left(1-\frac{v^2}{8f^2c_W^2}\right)$ 
\\
\hline 
\end{tabular}

\noindent
\begin{tabular}{|c|c|c|}
\hline
VVV & $\gamma W_H^+W_H^-$ & $ZW_H^+W_H^-$ \\
\hline\hline
$J$ & $-1$ & $c_W/s_W$
\\
\hline 
\end{tabular}

\noindent
The fields $\omega^\pm$, $\omega^0$ and $\eta$ are the Goldstone bosons 
of the [SU(2)$\times$U(1)]$_1\times$[SU(2)$\times$U(1)]$_2$ breaking into its diagonal subgroup. They are eaten by the heavy gauge bosons $W_H^\pm$, $Z_H$ and $A_H$, respectively. (Actually these Goldstone bosons mix with an additional physical Higgs triplet $\Phi$ at order $v^2/f^2$ and it is this linear combination of fields that is eaten.)
In principle, also the scalar triplet $\Phi$ contributes to the processes considered here. The corresponding diagrams can be obtained replacing $W_H^\pm$ by $\Phi^\pm$ and $Z_H,A_H$ by $\Phi^0$ and $\Phi^0_P$. The Feynman rules for the vertices containing $\Phi$, neglecting the masses of the SM fermions, involve couplings of ${\cal O}(v^2/f^2)$. As each diagram contains at least two such vertices, if any, they are suppressed by a factor of
${\cal O}(v^4/f^4)$ \cite{Blanke:2006eb}.

%%%%%%%%%%%%%%%%%%%%%%%%%%%%%%%%%%%%%%%%%%%%%%%%%%%%%%%%%%%%%%%%%%%%%%%%%%%%

\section{Loop integrals \label{apploop}}

%%%%%%%%%%%%%%%  FIGURE %%%%%%%%%%%%%%%%%%%%%%%%%
\begin{figure}[htb]
\centerline{\includegraphics[width=6cm]{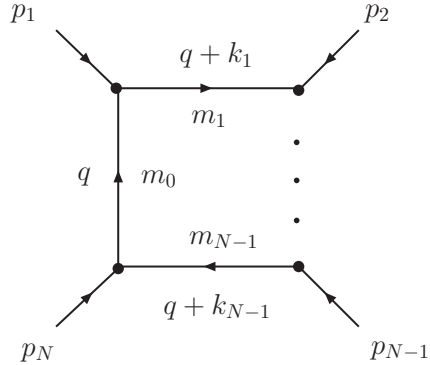}}
\caption{Generic one-loop diagram with $N$ legs.
\label{np}}
\end{figure}
%%%%%%%%%%%%%%%%%%%%%%%%%%%%%%%%%%%%%%%%%%%%%%%%%

Consider the generic one-loop diagram with $N$ legs in Fig.~\ref{np}, 
where
\bea
k_1=p_1, \quad k_2=p_1+p_2, \quad \dots \quad k_{N-1}=\sum_{i=1}^{N-1}p_i .
\eea
This diagram involves integrals of the type
\bea
\frac{\ii}{16\pi^2}T^N_{\mu_1\dots\mu_P}\equiv\mu^{4-D}\int\frac{{\rm d}^Dq}{(2\pi)^D}
\frac{q_{\mu_1}\cdots q_{\mu_P}} {(q^2-m_0^2)[(q+k_1)^2-m_1^2]\cdots[(q+k_{N-1})^2-m_{N-1}^2]}\ .
\eea
These integrals are symmetric under permutation of the Lorentz indices. The integration is performed in dimensional regularization. The mass scale $\mu$ keeps track of the correct dimension of the integral in $D=4-\epsilon$ spacetime dimensions.
$P\le N$ is the number of $q$'s in the numerator and determines the tensor structure of the integral (scalar for $P=0$, vector for $P=1$, etc.)
The notation is $A$ for $T^1$, $B$ for $T^2$, etc.
and the scalar integrals are $A_0$, $B_0$, etc.
The tensor integrals can be decomposed into a linear combination of Lorentz covariant tensors constructed from $g_{\mu\nu}$ and a linearly independent set of the momenta \cite{Passarino:1978jh}. The choice of the basis is not unique. Here we choose $g_{\mu\nu}$ 
and the momenta $k_i$, which are sums of the external momenta $p_i$ \cite{Denner:1991kt}. In this basis, the tensor-coefficient functions are totally symmetric in their indices. For this work, we need the following decompositions:
{\allowdisplaybreaks
\bea
B_\mu&=&
k_{1\mu}B_1\ ,
\label{t1}\\
C_\mu&=&
k_{1\mu}C_1+k_{2\mu}C_2\ ,
\label{t3}\\
C_{\mu\nu}&=&
g_{\mu\nu}C_{00}+\sum_{i,j=1}^2k_{i\mu}k_{j\nu}C_{ij}\ ,
\label{t4}\\
D_\mu&=&\sum_{i=1}^3 k_{i\mu} D_i\ ,
\\
D_{\mu\nu}&=&g_{\mu\nu}D_{00}+\sum_{i,j=1}^3 k_{i\mu}k_{j\nu} D_{ij} \ .
\eea}
These functions have been calculated for the argument configuration 
required by the processes under study, obtaining the following results.

\subsection{Two-point functions}

%%%%%%%%%%%%%%%  FIGURE %%%%%%%%%%%%%%%%%%%%%%%%%
\begin{figure}[htb]
\centerline{\includegraphics[width=6cm]{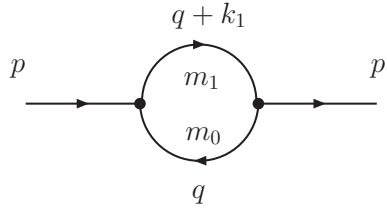}}
\caption{Generic one-loop diagram with two legs.
\label{2p}}
\end{figure}
%%%%%%%%%%%%%%%%%%%%%%%%%%%%%%%%%%%%%%%%%%%%%%%%%

Consider now the diagram with two legs in Fig.~\ref{2p}:  
\bea
\frac{\ii}{16\pi^2}\left\{B_0,\ B^\mu\right\}(\mbox{args}) 
=\mu^{4-D}\int\frac{\d^Dq}{(2\pi)^D}\frac{\left\{1,\ q^\mu\right\}}{\left(q^2-m_0^2\right) \left[(q+p)^2-m_1^2\right]}\ ,
\eea
where $k_1=p$. The corresponding tensor coefficients are functions of the invariant quantities $(\mbox{args}) = (p^2,m_0^2,m_1^2)$. The functions $B\equiv B(0;M_1^2,M_2^2)$ and $\overline B\equiv B(0;M_2^2,M_1^2)$ read
\bea
B_0=\overline B_0&=&\Delta_\epsilon+1-\frac{M_1^2\ln\dis\frac{M_1^2}{\mu^2} - M_2^2\ln\dis\frac{M_2^2}{\mu^2}}{M_1^2-M_2^2}\ , \\
B_1&=&-\frac{\Delta_\epsilon}{2}+\frac{4M_1^2M_2^2-3M_1^4-M_2^4+ 2M_1^4\ln\dis\frac{M_1^2}{\mu^2}+2M_2^2(M_2^2-2M_1^2)\ln\dis\frac{M_2^2}{\mu^2}}{4(M_1^2-M_2^2)^2} \nn\\
&=&-\overline B_0-\overline B_1\ ,
\eea
with $\Delta_\epsilon\equiv \dis\frac{2}{\epsilon}-\gamma+\ln4\pi$. These functions are ultraviolet divergent in $D=4$ dimensions.

\subsection{Three-point functions}

%%%%%%%%%%%%%%%  FIGURE %%%%%%%%%%%%%%%%%%%%%%%%%
\begin{figure}[htb]
\centerline{\includegraphics[width=5cm]{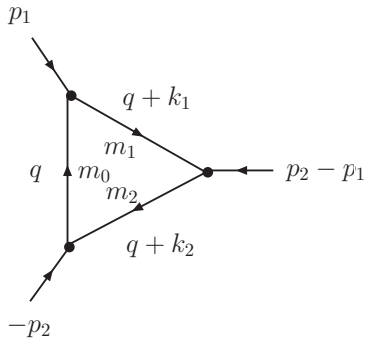}}
\caption{Generic one-loop diagram with three legs.
\label{3p}}
\end{figure}
%%%%%%%%%%%%%%%%%%%%%%%%%%%%%%%%%%%%%%%%%%%%%%%%%

Consider now the diagram with three legs in Fig.~\ref{3p}:
\bea
\frac{\ii}{16\pi^2}\left\{C_0,\ C^\mu,\ C^{\mu\nu}\right\}(\mbox{args})  
=\hspace{8cm}
\nn\\
\mu^{4-D}\int\frac{\d^Dq}{(2\pi)^D}\frac{\left\{1,\ q^\mu,\ q^{\mu}q^{\nu}\right\}}{\left(q^2-m_0^2\right) \left[(q+p_1)^2-m_1^2\right]\left[(q+p_2)^2-m_2^2\right]},
\eea
where we have chosen the extenal momenta so that $k_1=p_1$, $k_2=p_2$. The corresponding tensor coefficients depend on the invariant quantities
$(\mbox{args}) = (p_1^2,Q^2,p_2^2;m_0^2,m_1^2,m_2^2)$, with
$Q^2\equiv(p_2-p_1)^2$.
The functions $C\equiv C(0,Q^2,0;M_1^2,M_2^2,M_2^2)$ with $x\equiv M_2^2/M_1^2$ read
\bea
C_0&=&\frac{1}{M_1^2}\bigg[\frac{1-x+\ln x}{(1-x)^2}
\nn\\ &&\hskip1cm
+\frac{Q^2}{M_1^2}\frac{-2-3x+6x^2-x^3-6x\ln x} {12x(1-x)^4}\bigg]+{\cal O}(Q^4),\quad
\\
C_1=C_2&=&\frac{1}{M_1^2}\frac{-3+4x-x^2-2\ln x}{4(1-x)^3} +{\cal O}(Q^2),
\\
C_{11}=C_{22}=2\ C_{12}&=&\frac{1}{M_1^2}\frac{11-18x+9x^2-2x^3+6\ln x}{18(1-x)^4}+{\cal O}(Q^2),
\\
C_{00}&=&-\frac{1}{2}B_1
%\nn\\&&
-\frac{Q^2}{M_1^2}\frac{11-18x+9x^2-2x^3+6\ln x}{72(1-x)^4}
+{\cal O}(Q^4).
\eea
Or else, defining $\overline{C}\equiv C(0,Q^2,0;M_2^2,M_1^2,M_1^2)$,
\bea
\overline C_0&=&\frac{1}{M_1^2}\bigg[\frac{-1+x-x\ln x}{(1-x)^2}
\nn\\ &&\hskip1cm
+\frac{Q^2}{M_1^2}\frac{-1+6x-3x^2-2x^3+6x^2\ln x} {12(1-x)^4}\bigg]
+{\cal O}(Q^4),\quad
\\
\overline C_1=\overline C_2&=&\frac{1}{M_1^2}\frac{1-4x+3x^2-2x^2\ln x}{4(1-x)^3} ,
\\
\overline C_{11}=\overline C_{22}=2\ \overline C_{12}&=&\frac{1}{M_1^2}\frac{-2+9x-18x^2+11x^3-6x^3\ln x}{18(1-x)^4} ,
\\
\overline C_{00}&=&-\frac{1}{2}\overline B_1
%\nn\\&&
-\frac{Q^2}{M_1^2}\frac{-2+9x-18x^2+11x^3-6x^3\ln x}{72(1-x)^4}
+{\cal O}(Q^4).\quad\quad
\eea
Note that $C_{00}$ and $\overline C_{00}$ are ultraviolet divergent in $D=4$ dimensions.

In the limit $Q^2=0$ the following useful relations among two- and three-point functions hold:
\bea
\overline B_1+2\overline C_{00}&=&0,
\label{li1}
\\
-\frac{1}{4}+\frac{1}{2}\overline B_1+C_{00}-\frac{x}{2}M_1^2C_0&=&0,
\label{li2}
\\
-\frac{1}{2}+\overline B_1+6\overline C_{00}-xM_1^2\overline C_0&=&\Delta_\epsilon-\ln\frac{M_1^2}{\mu^2}.
\label{li3}
\eea

\subsection{Four-point functions}

The ones we need are all ultraviolet finite:
\bea
\frac{\ii}{16\pi^2}\left\{D_0,\ D^\mu,\ D^{\mu\nu}\right\}(\mbox{args})  
=\hspace{8cm}
\nn\\
\int\frac{\d^4q}{(2\pi)^4}\frac{\left\{1,\ q^\mu,\ q^{\mu}q^{\nu}\right\}}{\left(q^2-m_0^2\right) \left[(q+k_1)^2-m_1^2\right]\left[(q+k_2)^2-m_2^2\right]
\left[(q+k_3)^2-m_3^2\right]},
\eea
with $k_j=\dis\sum_{i=1}^{j} p_i$ and $\mbox{(args)}=(p_1^2,p_2^2,p_3^2,p_4^2,(p_1+p_2)^2,(p_2+p_3)^2;m_0^2,m_1^2,m_2^2,m_3^2)$. In the limit of zero external momenta, only the following integrals are relevant:
\bea
\frac{\ii}{16\pi^2}D_0&=&\int\frac{\d^4q}{(2\pi)^4}\frac{1} {\left(q^2-m_0^2\right)\left(q^2-m_1^2\right)\left(q^2-m_2^2\right) \left(q^2-m_3^2\right)},
\\
\frac{\ii}{16\pi^2}D_{00}&=&\frac{1}{4}\int\frac{\d^4q}{(2\pi)^4} \frac{q^2}{\left(q^2-m_0^2\right)\left(q^2-m_1^2\right)\left(q^2-m_2^2\right) \left(q^2-m_3^2\right)}.
\eea
In terms of the mass ratios $x=m_1^2/m_0^2$, $y=m_2^2/m_0^2$, $z=m_3^2/m_0^2$ the integrals above can be written as:
\bea
d_0(x,y,z)&\equiv& m_0^4 D_0 =
\left[
\frac{x\ln x}{(1-x)(x-y)(x-z)}-\frac{y\ln y}{(1-y)(x-y)(y-z)} \right.
\nn\\&&\hspace{2cm}\left.
+\frac{z\ln z}{(1-z)(x-z)(y-z)}
\right],
\\
\widetilde d_0(x,y,z)&\equiv& 4m_0^2  D_{00} =
\left[
\frac{x^2\ln x}{(1-x)(x-y)(x-z)}-\frac{y^2\ln y}{(1-y)(x-y)(y-z)} \right.
\nn\\&&\hspace{2cm}\left.
+\frac{z^2\ln z}{(1-z)(x-z)(y-z)}
\right].
\eea
For two equal masses ($m_0=m_3$) we get
\bea
d_0(x,y)&=&
-\left[
\frac{x\ln x}{(1-x)^2(x-y)}-\frac{y\ln y}{(1-y)^2(x-y)}+\frac{1}{(1-x)(1-y)}
\right],
\\
\widetilde d_0(x,y)&=&
-\left[
\frac{x^2\ln x}{(1-x)^2(x-y)}-\frac{y^2\ln y}{(1-y)^2(x-y)}+\frac{1}{(1-x)(1-y)}
\right].
\eea

%%%%%%%%%%%%%%%%%%%%%%%%%%%%%%%%%%%%%%%%%%%%%%%%%%%%%%%%%%%%%%%%%%%%%%%%%%%%

\end{document}